\documentclass[aip,graphicx,amsmath,amssymb,reprint,longbibliography,floatfix]{revtex4-1}

\usepackage{graphicx}
\usepackage{float}
\usepackage{color}

\begin{document}


\title{TEMPus VoLA: the Timed Epstein Multi-pressure Vessel at Low Accelerations} 



\author{H.L. Capelo}\affiliation{Space Research and Planetary Sciences Division, Physikalisches Institut, University of Bern, Sidlerstrasse 5, CH-3012 Bern, Switzerland}
\email{holly.capelo@space.unibe.ch}
\author{J. Kuhn}\affiliation{Space Research and Planetary Sciences Division, Physikalisches Institut, University of Bern, Sidlerstrasse 5, CH-3012 Bern, Switzerland}
\author{A.  Pommerol}\affiliation{Space Research and Planetary Sciences Division, Physikalisches Institut, University of Bern, Sidlerstrasse 5, CH-3012 Bern, Switzerland} 
\author{D. Piazza} \affiliation{Space Research and Planetary Sciences Division, Physikalisches Institut, University of Bern, Sidlerstrasse 5, CH-3012 Bern, Switzerland}
\author{M. Brandli}\affiliation{Space Research and Planetary Sciences Division, Physikalisches Institut, University of Bern, Sidlerstrasse 5, CH-3012 Bern, Switzerland}
\author{R. Cerubini} \affiliation{Space Research and Planetary Sciences Division, Physikalisches Institut, University of Bern, Sidlerstrasse 5, CH-3012 Bern, Switzerland}
\author{B. Jost}\affiliation{Space Research and Planetary Sciences Division, Physikalisches Institut, University of Bern, Sidlerstrasse 5, CH-3012 Bern, Switzerland}
\author{J.-D. Bod\'{e}nan}\affiliation{ETH Zurich, Institute of Geochemistry and Petrology, 8092 Zurich, Switzerland}
\author{T. Planchet}\affiliation{Space Research and Planetary Sciences Division, Physikalisches Institut, University of Bern, Sidlerstrasse 5, CH-3012 Bern, Switzerland}
\author{S. Spadaccia}\affiliation{Space Research and Planetary Sciences Division, Physikalisches Institut, University of Bern, Sidlerstrasse 5, CH-3012 Bern, Switzerland}
\author{R. Schraepler} \affiliation{Institut fur Geophysik und extraterrestrische Physik, Technische Universitat Braunschweig, Mendelssohnstr. 3, D-38106 Braunschweig, Germany}
\author{J. Blum} \affiliation{Institut fur Geophysik und extraterrestrische Physik, Technische Universitat Braunschweig, Mendelssohnstr. 3, D-38106 Braunschweig, Germany}
\author{M. Sch\"{o}nb\"{a}chler}\affiliation{ETH Zurich, Institute of Geochemistry and Petrology, 8092 Zurich, Switzerland}
\author{L. Mayer}\affiliation{1Center for Theoretical Astrophysics and Cosmology, Institute for Computational Science, University of Zurich, Winterthurerstrasse 190, CH-8057 Zurich,Switzerland}
\author{N. Thomas }\affiliation{Space Research and Planetary Sciences Division, Physikalisches Institut, University of Bern, Sidlerstrasse 5, CH-3012 Bern, Switzerland}

\date{\today}

\begin{abstract}

The field of planetary system formation relies extensively on our understanding of the aerodynamic interaction between gas and dust in protoplanetary disks. Of particular importance are the mechanisms triggering fluid instabilities and clumping of dust particles into aggregates, and their subsequent inclusion into planetesimals. We introduce the Timed Epstein Multi-pressure vessel at Low Accelerations (TEMPusVoLA), which is an experimental apparatus for the study of particle dynamics and rarefied gas under micro-gravity conditions. This facility contains three experiments dedicated to studying aerodynamic processes, i) the development of pressure gradients due to collective particle-gas interaction, ii) the drag coefficients of dust aggregates with variable particle-gas velocity, iii) the effect of dust on the profile of a shear flow and resultant onset of turbulence. The approach is innovative with respect to previous experiments because we access an untouched parameter space in terms of dust particle packing fraction, and Knudsen, Stokes, and Reynolds numbers. The mechanisms investigated are also relevant for our understanding of the emission of dust from active surfaces such as cometary nuclei and new experimental data will help interpreting previous datasets (Rosetta) and prepare future spacecraft observations (Comet Interceptor). We report on the performance of the experiments, which has been tested over the course of multiple flight campaigns. The project is now ready to benefit from additional flight campaigns, to cover a wide parameter space. The outcome will be a comprehensive framework to test models and numerical recipes for studying collective dust particle aerodynamics under space-like conditions.

\end{abstract}


\maketitle 

\section{Introduction}
The exploding field of exoplanetary science has revitalized inquiries into the origins of
planetary systems. Understanding planetary system formation relies extensively on physical
models and numerical simulations to explain the formation of planets from the gas and dust in
protoplanetary disks. Observational constraints for formation theories are available in the
Solar System in the form of chondritic meteorites\citep{Scott2014} and comets\citep{ZolenskyEtal:2006} that are thought to be remnants
of the material from which planets formed in the inner and outer Solar System, respectively\citep{Nuth2018}.
Progress in imaging protoplanetary disks in various phases of their evolution\citep{ALMA2015} is also helping to inform the models on the planetary formation processes. However, it is also crucial to test and verify experimentally the
key aspects of the physics underlying the computational models. Of particular importance for
planetary formation are the mechanisms triggering clumping of dust particles into aggregates.
Some of the resulting aggregates might eventually become chondrules\citep{Bodenan2020} or igneous Calcium-
Aluminium-rich Inclusions (CAIs)\citep{Macpherson1993}, following a melting event, while icy dust aggregates at
larger heliocentric distances might still be preserved in primordial comets. These mm- to cm sized
aggregates, often referred to as pebbles, might also be key in explaining how to
assemble km-sized planetesimals from micrometer-size dust particles, overcoming the
“metre-size barrier" problem\citep{Weidenschilling:1977a}. Often, these mechanisms involve instabilities arising from the interaction between the gas and the dust phase in the disk, and, specifically, are driven by drag forces acting between gas and dust.

At nearly all evolutionary stages of the protoplanetary disk, the hydrodynamic interaction
between dust and gas mediates the transport and growth of planetary precursors. Of particular
importance is aerodynamic drag, not only because it enters into the drift dynamics of small
bodies \citep{BrauerEtal:2008a}, but also because particles' collective back reaction to drag forces
alter the state of the bulk fluid properties, and consequently the propensity to become
unstable. This is true for heavily mass-loaded fluids in which the particles are poorly coupled,
which result in streaming instabilities \citep{You_good2005}, but also at the interface
between particle-rich and particle-poor flows, where shear instabilities have been
hypothesised to develop \citep{Bai_2010b,Surville:2018}. Recent theoretical and experimental results of multi-particle sedimentation in rarified gas show evidence for a
localised pressure gradient induced by swarms of particles \citep{lambrechts,Capelo:2019}, with dependency on the spatially averaged dust-to-gas mass density ratio. Such a
mechanism is thought to underlie, for example, the streaming instability. Moreover, \citep{Lin:2017} showed that the set of equations describing such behaviour can be generalised
and applied to various instabilities in disks, by adopting a modified equation of state for
particle-laden flow. We identify the question of how local pressure gradients can develop in
granular material with extremely diffuse particle-seeding density as an important process to
formally investigate. Thus far, the experiments presented in \citep{Capelo:2019} are the only
instance in which a dust-drag fluid instability has been generated in a laboratory flow with
scale similarity to the flow in a protoplanetary disk. The mechanistic explanation for this
effect calls upon simple Newtonian physics, whereby particles acted upon by the gas, act
back, and in so doing slow the gas motions locally. Notably, the same basic mechanism is at
play in driving the instability simulated by \cite{Surville:2018}, and so we also generically group the complex dynamics studied there as a dust-drag instability.

Chondritic meteorites and comets are generally considered to hold the most direct records of
the early Solar system conditions and hence the planetesimal formation process. Amongst the
many important findings of the Rosetta Mission, was the confirmation that the particles
ejected from the comets nucleus were in fact irregular-shaped aggregates composed of submicrometer
grains\citep{Mannel:2016}. Previous experiments have already determined the
drag-force coefficient of restitution in the Epstein drag regime in micro-gravity \citep{BlumEtal:1996} but without considering the relative velocity between gas and particles. Moreover, the effect of porosity on aggregate drag force in the Epstein regime has not been addressed
experimentally. In particular, in modelling two-fluid instabilities, dust particles are treated as smooth spherical bodies, while it is expected that they would be irregular in shape and porous,
based on evidence from meteorites and comets. The drag force regime would be different in the latter case compared to standard assumptions. Addressing this is novel, and is of fundamental and general importance.

Synergies between theory, experiment, and space-based observation lend support to the
scenario that planetesimals are formed by aero-gravitational instability, largely evidenced by
the extremely high porosity \citep{mohtashim,Wahlberg:2014,blum:2014} and low tensile strength of 67P \citep{Attree:2018}. However, the models leading to
such conclusions often hold strong assumptions regarding the shear and tensile strengths of
the materials involved and generally follow the tradition of treating all dust as silica. This
assumption is obviously too simplistic since there is plentiful evidence from the condensation
sequence of dust in circumstellar environments and from the cosmochemical record that other
minerals and organic molecules are also abundant \citep{YONEDA:1995,Petaev:2005}. One of our important objectives is to address how the use of more realistic dust analog materials may impact the observable physical properties of planetesimals. In particular, we can address the compactness of different granular materials when subject to variations in differential gas pressure and gravitational load, which is analagous to outgassing and collisions between parent bodies, respectively. 

While the project described in this paper focuses on topics related to planetary formation,
we also note that many of the physical mechanisms investigated in this context are also
relevant for the evolution of Solar System objects. This is particularly true for cometary nuclei
where low-pressure streams of gas produced by the sublimation of the ice interact with the
dust, lifting it from the surface and accelerating it away from the nucleus. Some of the results are therefore also applicable to the interpretation of data from
instruments on past cometary missions (Giotto and Rosetta in particular) and helpful to plan
future investigations (Comet Interceptor). 

The purpose of the TEMPus VoLA facility is to unify the description of collective dust-grain
aerodynamics (drag, back reaction, effect on flow viscosity, and symmetry breaking) by
spanning from the classically studied regime (high pressure, high packing density of particles)
to approach the flow conditions expected in protoplanetary discs (extremely dilute gas, very
low particle seeding density) and some Solar System objects.

The need for microgravity experiments is due to the general difficulty of fluidizing (levitating) inertial particles in a low-density gas. Moreover, in granular beds, the porosity or equivalently filling factor, is determined by gravitational sedimentation. To represent the low filling factors of particles on the surface of a low-gravity body (e.g. flows on and around asteroidal or cometary objects, or loosely bound dust aggregates) one requires reduced-gravity conditions.  A facility dedicated to levitate dust grains lends itself to the study of their radiative transfer properties as well \citep{PROGRA2}.

The most important goal leading the experimental design principles is to satisfy a set of dimensionless fluid-dynamical parameters, which can be scaled to the conditions in space. In particular we consider:  i) the Knudsen number of the flow (ratio of the gas mean-free path to particle or pore size); ii) the mean inter-particle distance or porosity of the particle suspension; iii) the mean dust-to gas mass density ratio; iv) the Reynolds number; v) and the particle stopping time or, equivalently, viscous relaxation length scale. Recall the principle of dynamic similarity: any flow will be similar to another flow, if they have the same non-dimensional numbers. Therefore, fluid dynamics experiments can be a very powerful tool to predict the behaviour of particle-rich fluids that exist in space, which are often simulated, but otherwise difficult to study by any direct empirical means. 
Such experiments are the topic of this work. We summarise the three separate experiments of the TEMPus VoLA facility, with which we explore three limiting regimes:  
\begin{itemize}
\item {{\bf Experiment one} Gas permeability of free-molecular flow through granular media with porosity 50-99\%, which are much higher than can be achieved on Earth and comparable to planetesimal porosity;}
 \item{{\bf Experiment two} The dependence of Epstein drag on aggregate porosity, when subjected to variable gravitational load and therefore variable relative dust-gas velocity;} 
\item{{\bf Experiment three} the two-fluid limit in which the coupling between the gas and dust is mediated only by the exchange of momentum for tightly coupled particles and for which, in the absence of gravity, a shear profile--and potentially shear instability--is predicted to develop.}
\end{itemize}

For the most part, these processes are well-studied theoretically, and less-often addressed experimentally. Moreover, previous experiments have never satisfied the complete parameter regime to which we strive. 

This document is organised as follows: In section \ref{principles} we first introduce the set of equations and parameters that define our systems. We then establish the uniqueness of our parameter space as well as the similarity to flow conditions in the planet-formation scenario. We then outline the concrete set of hypothesis that can be tested experimentally. In section \ref{Design} we present the technical details of the experiments and in \ref{verification} we report on the two-phase flow conditions obtained.   
In section \ref{future} we summarise the project and delineate future directions for the facility.  

\section{Principles and Parameters}\label{principles}

\subsection{Fluid equations}
When the length scale, $L$, of a gaseous volume under consideration is much larger than mean inter-molecular separation, $\lambda$, the gas dynamics can be described as a continuum, with the Navier-Stokes equations expressing the balance of mass and momentum. In the incompressible limit, mass is conserved,

\begin{equation}
\nabla \cdot \vec{U}=0 
\end{equation}
and momentum balance is given by,

\begin{equation}\label{ns_general}
\rho \left[ \frac{\partial u}{\partial t} + (\mathbf{u}\ \cdot \nabla)\mathbf{u} \right] = \vec{f} - \nabla P + \nu \nabla^{2}\mathbf{u}.
\end{equation}

The kinematic viscosity, $\eta=\nu/\rho$, is the ratio of the molecular viscosity to the density of the fluid. In interplanetary space, the gas density is very low, of the order $\rho \sim$10$^{-9}$ g cm$^{-3}$ or less, and so the viscosity is proportionately driven up and determined mainly by the density of the gas, for any atomic or molecular gas species. For this reason, it is highly justified to perform experiments with a common gas such as air, and control the gas density in order to specify the correct order of magnitude of the quantity $\eta$. 

In a rotating accretion disc, the body forces include the Coriolis, centripetal and centrifugal forces. The latter two are conservative and therefore can be considered as effective pressure (moved into the absolute pressure term), whereas the Coriolis force cannot. Rotating discs are therefore not Gallilean invariant to frame transformations which is a pre-requisite to evoke scaling relationships between self-similar flows; in Newtonian physics, the Galilean invariance ensures that the conservation laws of fluid motions remain the same in any two reference frames--such as rotating and non-rotating. For a full proof of the non-invariance of rotating flows with non-negligible Coriolis force, see for example \cite{Pope:2011}. The Rossby number, Ro, is a dimensionless number describing the relative importance of global rotation with respect to local inertial circulation: 

\begin{equation}
Ro=\frac{U}{ L\left| \omega \right| }\sim \frac{\vec{u} \cdot \nabla \vec{u}}{\Omega \times \vec{u}}.
\end{equation}

When L is taken to be the disc scale height, Ro becomes low and so Coriolis forces cannot be neglected. However, deriving the intrinsic properties of a particle-laden fluid may be addressed on the local, rather than global, scale. For a typical particle of mass $m$, and drag force $F_{\rm d}$ the dust stopping time scale is $T_{f}=m|u|/F_{d}$. \cite{lambrechts} showed that for a typical disc model, the associated stopping length scale of an inertial particle is coincident with the high $Ro$ regime, so that  experiments involving only conservative forces could address the fluid in question. We adopt the same approach here, avoiding rotation.  For such a conservative system transformed into Earth's reference frame, the only body force is a constant gravitational acceleration, $g$. 

Further simplifications to equation \ref{ns_general} result when the typical magnitude of convective accelerations is small with respect to the viscous stress. This criterion is expressed by the Reynolds number: 
\begin{equation}
Re =\frac{UL}{\nu} \sim \frac{\left| \rho(\mathbf{u} \cdot \nabla)\mathbf{u}\right|}{\left| \eta \nabla^{2}\mathbf{u}\right|}.
\end{equation}

Note that when the disc scale height determines L, then $Re\rightarrow \inf$ and the fluid is considered to be inviscid. However, if the diameter of dust grains sets the value of L, then viscosity cannot be ignored. 

For a single particle in a fluid, a Stokes number, comparing the importance of the local acceleration to the viscous term can be defined:

\begin{equation}
St=\frac {a^{2}}{T\nu}\sim \frac{\left|\rho \partial u /\partial t \right|}{\left| \eta \nabla^{2}\mathbf{u}\right|},
\end{equation}

Where $a$ is the mean inter-particle separation. The requirement that  $St<1$ leads to a momentum diffusion time scale $t_{d}=n^{-2/3}/\eta$. Provided that $t_{d}> T_{f}$ one assumes that the particles do not perturb one another via viscous interactions. This criterion is important because such condition allows one to model a two-phase flow using a so-called `pressureless fluid' dust model \citep{You_good2005}. It was established in \cite{lambrechts} that such criterion applies directly in the context of protoplanetary discs, since the particle seeding density is very low, the particles do not frequently interact, and so can be treated collectively as a fluid. In this case, the momentum equations are a set of coupled equations which feed-back via drag force. In the absence of external forces, the gas velocity is expressed as

\begin{equation}\label{vgas_equation}
\partial v_{gas}/\partial t= \frac{\epsilon}{t_{f}}[v_{dust}-v_{gas}]
\end{equation}
and the particle velocity is expressed as
\begin{equation}\label{vdust_equation}
\partial v_{dust}/\partial t = - \frac{1}{t_{f}}[v_{dust}-v_{gas}]
\end{equation} 

While these equations are time dependent, they reach an equilibrium value with a relative velocity which is strongly determined by the two parameters $\epsilon$, the dust to gas density ratio, and $t_{f}$. It has been demonstrated both theoretically and experimentally, that values of $\epsilon$ near 1 or greater result in a streaming instability \citep{Capelo:2019}. However, we see that even for small, but non-negligible, values of $\epsilon$, a relative dust-gas velocity can occur. This is a critical matter, since the resulting shear may affect the bulk fluid dynamics.  The stopping time is also important, since it will also reduce the relative velocity, and for extremely long stopping times, the coupling between the phases becomes weak. 

We note that it is conventional to cite a Stokes number in the context of protoplantary discs \citep{Weidenschilling:1977a, Sekiya1988,Nakagawa, bai_stone1}, which is a comparison of the orbital time scale to T$_{\rm f}$. The meaning of this comparison is to establish how quickly particles react to changes in acceleration of the fluid, and in discs, the rate of particle drift is crucial for the timescales over which particles are available to help form planets.   The Stokes number defined as such is in general a proxy for particle size, since the relative gas-dust headwind velocity is a constant throughout the disc, and since the density of the particles is generally taken to be constant. The particle size has implications for how particles move in the disc, and in particular it determines the drag law that applies.

As initial conditions for planet formation, one assumes that the initial density and temperature profiles of the dust and gas in a protoplanetary disc are axis-symmetric in the azimuthal direction and follow a power-law decrease in the radial direction away from the star. The scalings for pressure and temperature can then be used to calculate the mean free path of the gas,

\begin{equation}
    \lambda=\frac{1}{n\sigma}=22{\rm R}^{9/4}_{\rm AU}{\rm cm}
\end{equation}
R is the heliocentric distance of Earth, in Astronomical Units (AU). Assuming the gas is primarily molecular hydrogen with  molecular radius 10$^{-8}$ cm, the cross-section, $\sigma=2\times 10^{-16}$ cm$^{-2}$. The number density at the midplane of the disc, n$=1.5 \times 10^{14}$ R$^{-9/4}$ g cm$^{-3}$ comes from the following relations for surface density, temperature, and angular velocity:
\begin{equation}\label{mfp}
    \Sigma= 300~R^{-1}_{\rm AU}{\rm ~g~cm}^{-2}
\end{equation}
\begin{equation}
    \\T=280~{\rm R}^{-1}_{\rm AU}~{\rm K} 
\end{equation}
\begin{equation}
    \Omega=2\times 10^{-7} {\rm R}^{-1/2}_{\rm AU}{\rm s}^{-1}
\end{equation}
               v Since the sound speed is defined as $c_{s}= \sqrt{P/\rho}$, and the vertical scale height $H=c_{s}/\Omega$, both the mid-plane gas mass and number density result.

\begin{center}
\begin{figure}[ht]
\includegraphics[width=0.5\textwidth]{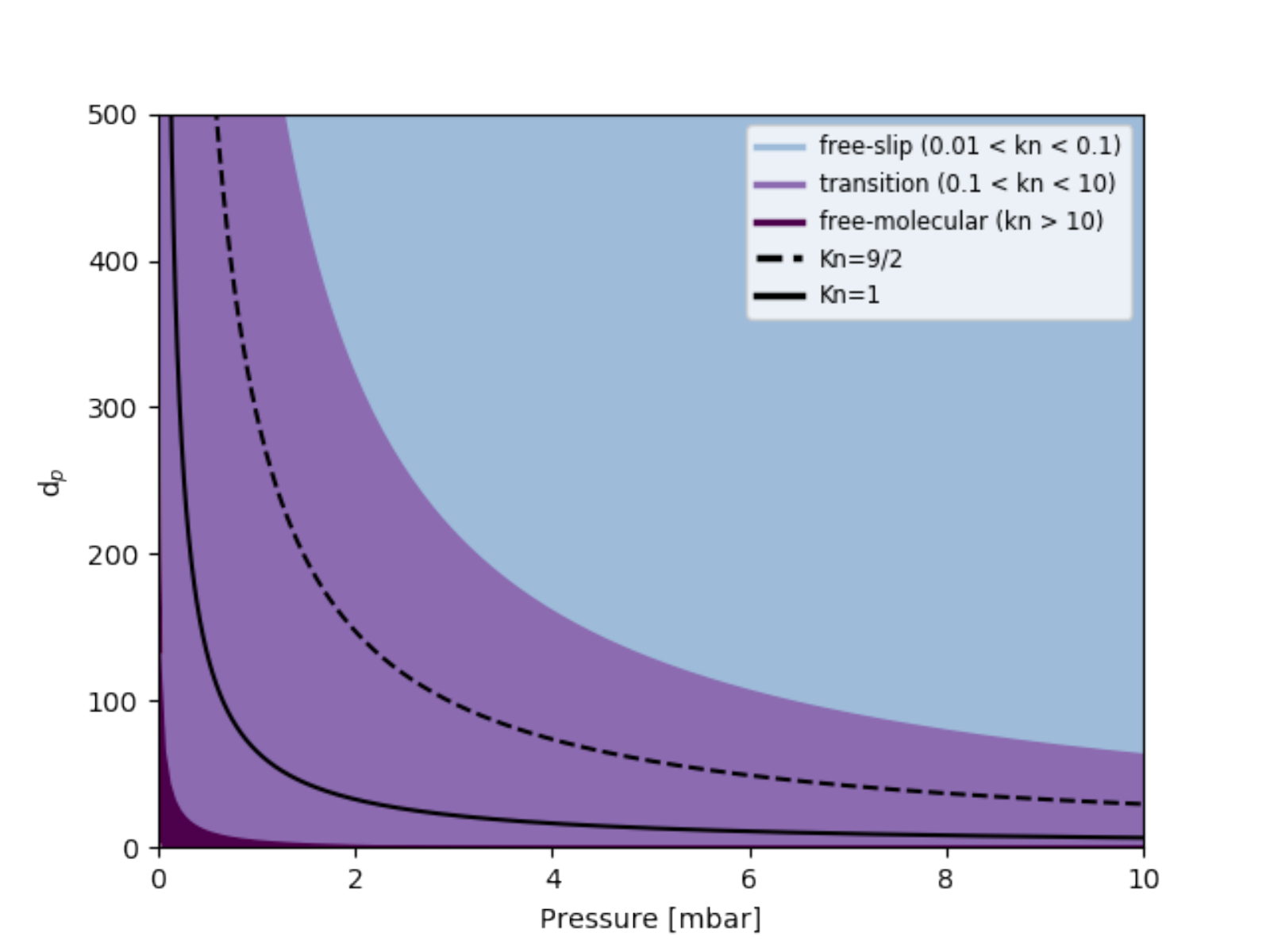}\\
\caption{\label{lab_knudsen} Comparison of particle size to mean inter-molecular distance in a laboratory setting. Solid black line: length scale (mean free path $\lambda$, particle size) as a function of gas pressure (air, room temperature). Drag regime divisions are indicated in the legend, with light blue corresponding to the free slip regime, transitional flow in violet, and free molecular flow in plum. Note that these divisions are subject to definition and are gradual. The pure continuum flow regime is not relevant at these values of Kn. The aerodynamic drag regime experienced by particles of meter scale in protoplanetary discs (always in transitional flow or free molecular flow) can be accessed by studying particles in the size ranges of a few tens hundreds of micrometres and gas pressures below 10 mbar.}%
\end{figure}
\end{center}

The dimensionless number Kn delineates the aerodynamic drag regime which applies.  The transition from `Stokes' to `Epstein' Drag regime is often set at the value of Kn$=9/2$. The value 10$<$Kn is conservative, some sources \cite{Allen_Raabe:1985} claim free molecular drag applies as low as Kn$\sim~2-3$, and in fact the turnover in drag law, when using the Cunningham correction, is steep around Kn$\sim$1. According to equation \ref{mfp}, the value of $\lambda$ is 22 cm at the disc miplane at 1 AU, and only increases at greater distances. Therefore, dust particles nearly always have Kn $\ge$ 1, and so fall into the Epstein drag regime, and it is only at particles sizes of a few decimeters or greater that the flow conditions can be represented by the Stokes drag law.  We notice that for the vast majority of the disc, all objects smaller than km-scale are in the transition or Epstein drag regime. 
In Figure \ref{lab_knudsen}, we demonstrate at which experimental gas pressures one can achieve the same values of Kn, as are found in the planet-formation scenario. We show different values of Kn, corresponding to a choice of air pressure and particle size.  For example, the Kn$\sim1$ regime is obtained with gas pressures of $\sim1$ mbar and particles of $\sim100$ $\mu$m. 
It is noteworthy, that in some contexts, the particle concentration becomes high, and gas must flow through a porous matrix. An example of the momentum equations \citep{2007PhRvL..99d8001V} for the gas, and particles with mass m, respectively, are then:
\begin{equation}
\phi \left( \frac{\partial P}{\partial t} + \vec{u} \cdot \nabla P \right)=\nabla \cdot \left(P \frac{\kappa(\phi)}{\mu} \nabla P \right) - P\nabla \cdot \vec{u}
\end{equation}
\begin{equation}
m \frac{d\vec{v}}{dt}=mg+F_{l}-\frac{V\nabla P}{\rho}
\end{equation}
Where P is the pressure, $\phi$ is the particle filling factor. The quantity $\kappa(\phi)$ is of particular interest, as it derives from the simpler, classic example of Darcy flow, and has the same meaning as the permittivity coefficient \citep{darcy_early}. Recent studies have revisited Darcy in the slip regime \citep{maria}, with applications to pre-planetary bodies. We note however, that such small bodies also have low gravity and so the packing of the granular material is lower than it is due to Earth's gravity. Consequently, it is of interest to investigate flow through granular media at intermediate and low packing fractions afforded by zero or partial gravity platforms.  Moreover, the principle of a clustered granular medium applies to dust aggregates themselves, even though they are generally modeled in a highly simplistic manner, such as by assuming them to be solid spherical `pebbles'.  
\subsection{Drag laws}
For an individual object embedded in the flow (low $\phi$), the resistance due to relative motion with the gas that it experiences can be calculated by integrating the stress tensor $\sigma$ over the surface area. 

While we stated above that gas density is the main determinant of viscosity, there remains the question of how the existence of an embedded particle population modifies the fluid viscosity itself. 
 
For flow in a constant mean direction through a granular bed (high $\phi$), the difference in pressure at a given height is moderated by the average pore opening area $\kappa \propto1-\phi$:
\begin{equation}
\frac{dP}{dh}=-\frac{\eta}{\kappa}\vec{u}.
\end{equation}

\cite{Brinkman} was the first to recognise the similarity of this expression to a drag force, and showed that the resulting integration of the stress over the surface of a sphere results in the following drag law:

\begin{equation}\label{Brinkman}
F_{d,Brinkman}=6\pi \nu' u R_{p}[1+\lambda R+ \lambda^{2}R^{2}/3].
\end{equation}
Here R is half the particle diameter $d_{p}/2$, and $\lambda=(\frac{\nu}{k\nu'})^{1/2}$. The viscosity of the particle-laden flow, $\nu'$, is the fluid viscosity modified by a factor $k(\kappa)$. When $k=1$, $\nu'$ can be replaced by $\nu$ and equation \ref{Brinkman} returns the Stokes drag formula.  The form and derivation of $k$ have been debated since first proposed by \cite{einstein_viscosity:1905}, however there is plentiful evidence that drag is reduced for dense swarms of particles in the continuum flow regime. Whether this relation holds at all in the free molecular regime has not yet been addressed either theoretically or experimentally.

By a similar reasoning, fluid drag on porous particle aggregates, often treated as spheres of equivalent surface area, ought to be subject to a correction dependent upon the permittivity ($\kappa_{material}$), of the material in question.  Consider the potential importance of this possibility by noting that in the Epstein regime the drag coefficient $f$ is determined by the probability $\chi$ of a specular or diffuse reflection of molecules from the particle's surface 

\begin{equation}
f=\chi f_{sp}+(1-\chi)f_{di}.
\end{equation}

for specular reflection 
\begin{equation}
f_{sp}=\frac{4}{3} \pi d_{p}^{2} \rho_{g}v_{th},
\end{equation}
with $v_{th}$ being the thermal velocity of the gas. For diffuse reflection
\begin{equation}
f_{di}=f_{sp}(1+\pi/8).
\end{equation}

We see that this formulation has no allowance for more complex processes such as flow through the porous structure. We therefore investigate this topic experimentally.

\section{Rationale and aims of the experiments}
The study of fluid dynamics rests on the foundational principles of Navier-Stokes, which is that a flowing fluid can be described by mass and momentum continuity equations. In rarefied gas, the control parameters setting the behavior of the fluid described by such equations should depend upon Knudsen number Kn, dust-to-gas density ratio $\epsilon$, particle packing fraction $\phi$, the Stokes St (equivalently drag coefficient), and Reynolds Re numbers. 

The TEMPusVoLA facility is divided into three systems, so that we can study limiting cases and disentangle the relative importance of these parameters, specifically for values that match those expected in the planetesimal formation and evolution scenario. We state the expectations associated with each system separately.  

System 1 Permeability:  

Premises:  The permeability of a porous medium to gas diffusion is a function of Knudsen number and particle packing fraction; the role of inertia is usually parameterised in permeability studies via the Reynold’s number;  At low gas pressure, the Reynolds number effect is considered negligible due to the dependency on gas density; when density approaches zero, Reynold's number approaches zero, because Re$=\rho v L/\mu$, where the numerator is the product of the gas density, velocity and typical length scale and the denominator is the molecular viscosity of the gas.

Hypothesis: \cite{Lin:2017} predicted that even for a flow in which: i) Re is low due to vacuum gas pressures, and ii) $\phi$ is low due to lack of gravitational compression, there should be an effect on the pressure gradient across a porous medium, due to the collective inertia (parameterised by~$\epsilon$) of the suspended particles. 

System 2 Stokes Numbers of porous aggregates:  

Premises: Drifting pebbles that participate in fluid instabilities in protoplanetary discs necessarily have a high relative particle-gas velocity and their drag coefficient depends upon Knudsen number (Kn$=\lambda$/L, the ratio of the gas molecules’ mean free path and the size of the particle); the pebbles in a protoplanetary disc are not solid spheres, but rather porous aggregates and so gas can flow not only around but through them.  

Hypothesis: The slip drag coefficients for porous aggregates should depend on both Knudsen number and the permeability coefficient of the porous aggregate. 

System 3 particle-laden shear flow:
  
Premises: When two fluids of different density flow next to one another and are acted upon by a force, the difference in force leads to a difference in velocity and causes a shear profile to develop. 

Hypothesis: If the high-density layer of a flow is seeded with particles, and in the absence of all other external forces, the collective forcing of the particles on the gas will cause a slowing of the gas and result in a developing shear profile. For specific values of $\phi$ and $\epsilon$, this profile can become unstable and cause turbulence.

\section{Design of the instrument}\label{Design}

\subsection{Systems overview and integrated racks}
 We break down the TEMPus VoLA facility into six ``systems'':\\
\begin{itemize}
 \item{Systems 1-3: correspond to the three experiments: 1. Permeability chamber, 2. Drag chamber, and 3. Shear flow chamber} 
\item{System 4: High-speed camera acquisition \& storage}
\item{ System 5: Vacuum (pump, valves, pressure sensors, data logging, mass flow control)}
\item{System 6: monitoring acquisition \& storage}

\end{itemize}

\begin{center}
\begin{figure}

 \includegraphics[width=0.5\textwidth]{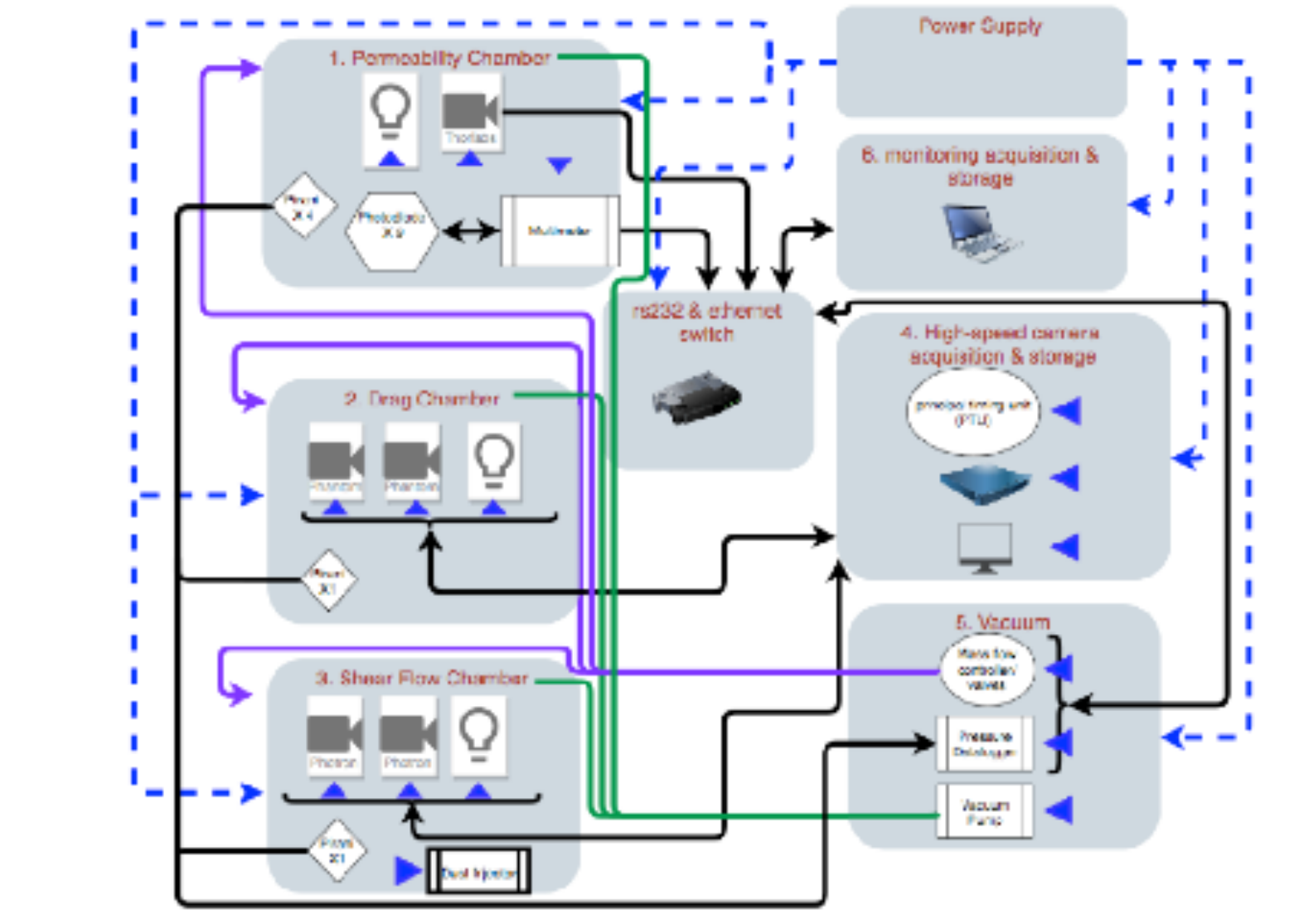}
 \caption{\label{all_systems} The six main systems of the TEMPus VoLA facility. Black lines represent data transfer, green lines represent vacuum tubes, purple lines are part of the gas intake line, blue dashed lines are power cables. Hardware items requiring a power outlet are marked with blue triangles. The permeability chamber (system 1) and vacuum system (System 5) are controlled and monitored by a laptop (system 6). The drag chamber (System 2) and shear flow chamber (system 3) involve recordings by high-speed cameras which are controlled and synchronised with a pulsed light source by the high-speed acquisition and storage (system 4).}%
 \end{figure}
 \end{center}

Fig. \ref{all_systems} provides a schematic representation of the measurement and control equipment related to each of the experiments. 
Note that System 4 is common to both Experiments 2 and 3 because they both require the use of high-speed cameras and synchronised pulsed lighting. System 5 is common to all three of the experiments, which share a vacuum line, vacuum flow controller, and pressure datalogging system. System 6 refers to the laptop used to collect images of Experiment 1 during its operation. 
\onecolumngrid
 \twocolumngrid

The three experiments are fixed on a primary rack, which is optimised to meet the loading tolerances of parabolic flight. The specific dimensions and detailed design features can be inspected in the mechanical drawings supplied in the appendicies \ref{append}.  In particular, the center of mass and bending radius was observed in the placement of the apparatus onto the primary rack. The configuration of the three experiments was decided with consideration for the drift in acceleration throughout the parabolas, which varies depending upon the axis of the plane, and is most pronounced along the cockpit-to-tail direction. Below we will comment upon the relative sensitivity of our measurements to this drift and the ways that we address it. The primary rack also involves a removable external casing, which is a safety requirement intended to contain stray aerosols or broken equipment, in the unlikely event of a system failure involving breakage. 

\begin{center}
\begin{figure}

\includegraphics[width=0.5\textwidth]{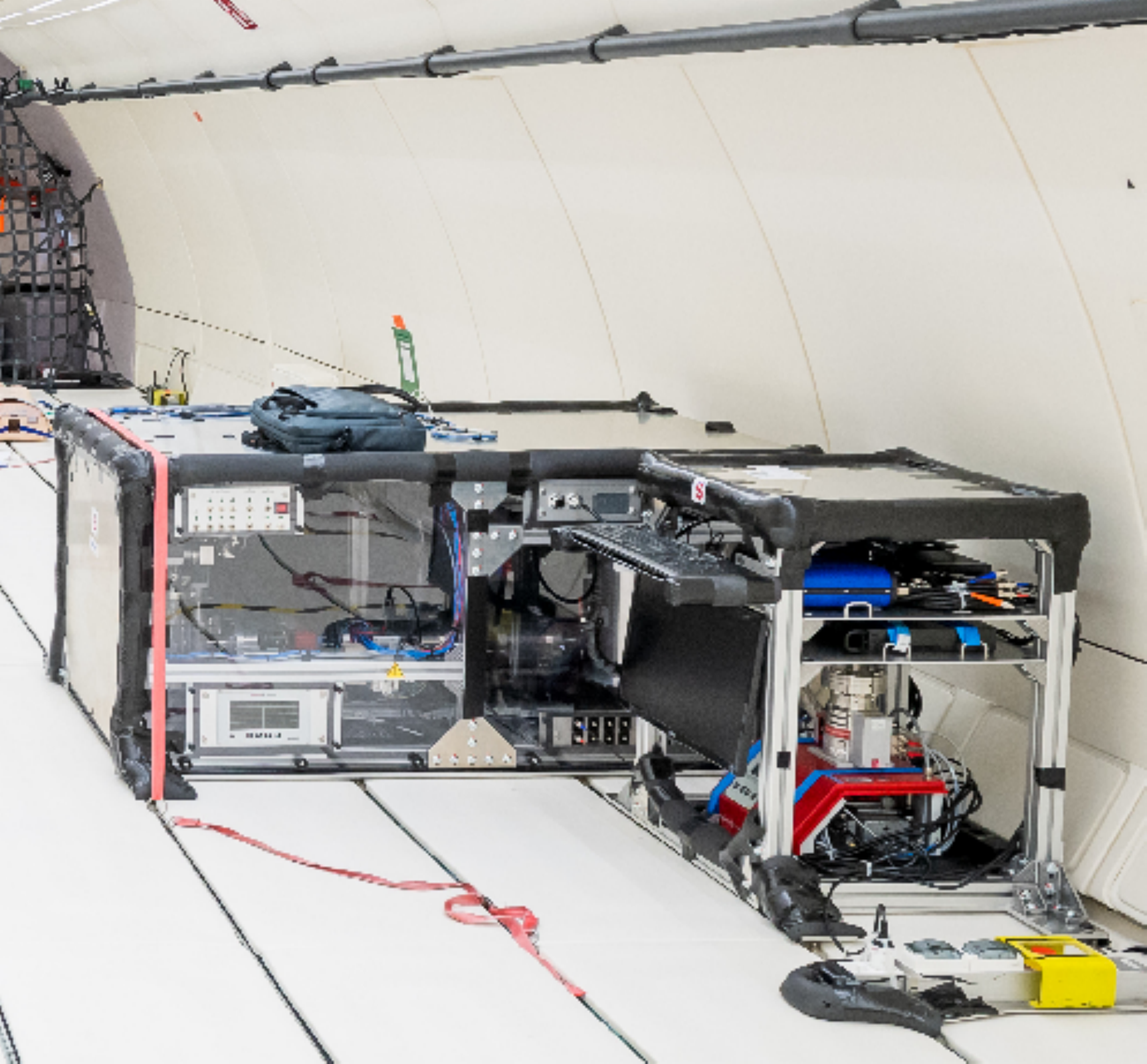}\\
\includegraphics[width=0.5\textwidth]{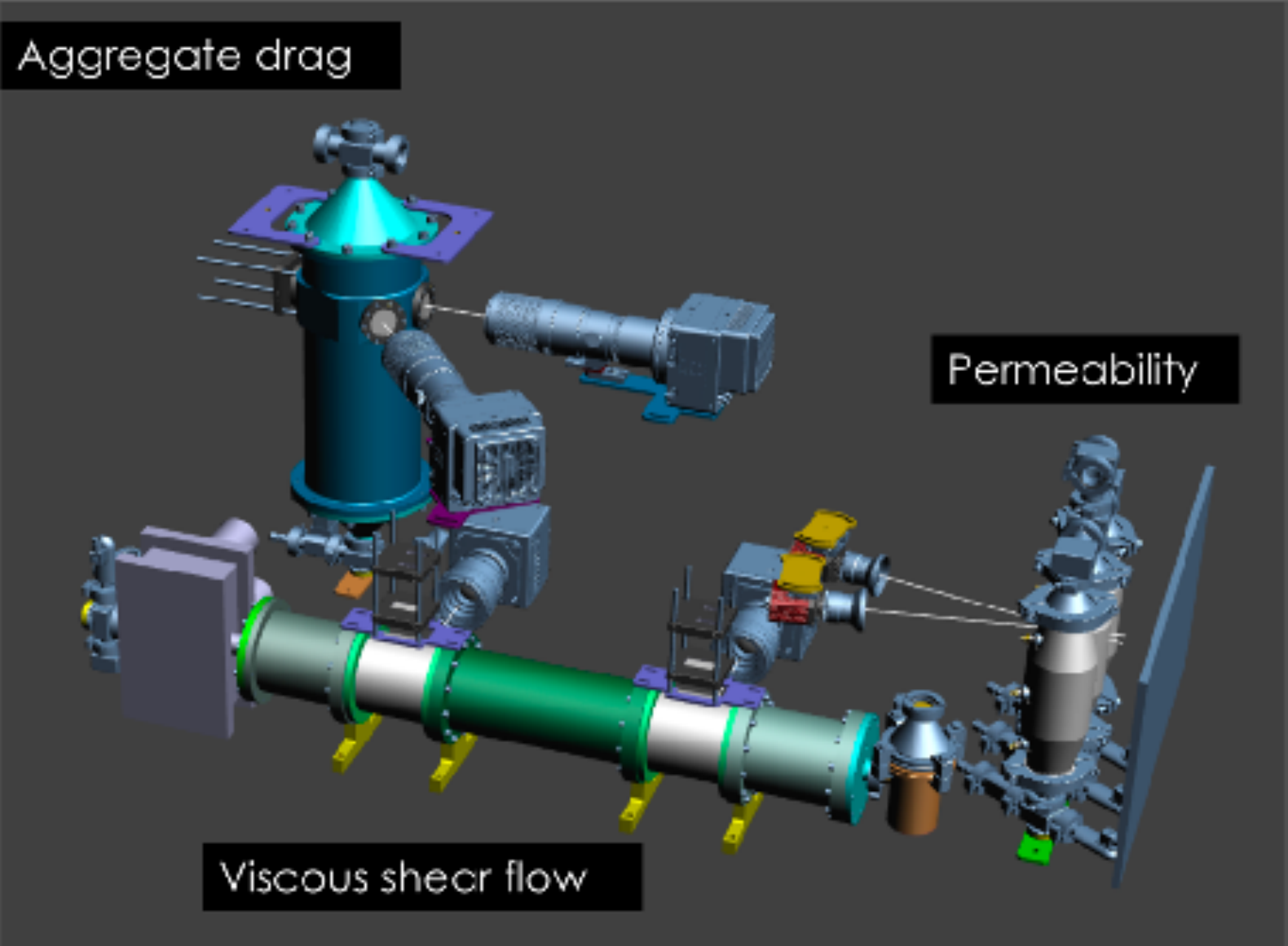}\\

\caption{\label{racks}{\bf Top}: Both primary (larger) and secondary (smaller) rack in flight configuration aboard Air Zero-g. The laptop placed on the primary rack controls the camera related to system 1 and the mass flow controller (MFC). All three experiments are connected to the same vacuum line. The top shelf of the secondary rack holds two pulsed light sources and timing unit, the second shelf holds the computer server and the bottom shelf holds the vacuum pump. The computer on the secondary rack controls the high-speed cameras, timing unit, and pulsed light source belonging to the high-speed three-dimensional particle-tracking system. The switch panel to control the vacuum valves, ventilation fan, dust injector, and LED panel is shown on the top left of the primary rack. The optical-fiber and vacuum-tube feedthroughs are shown connecting the two racks. {\bf Bottom}: The experimental apparatus, shown with neither external case nor connecting cables and tubing. To the right of the image is experiment one (three clear vessels with backlighting), to the left of the image is experiment two (tall vertical cylinder with viewing ports), and experiment three is towards the front of the configuration (long horizontal chamber with clear segments).}%
\end{figure}
\end{center}

Several additional components are held on a secondary rack, including the vacuum pump, light sources for the high-speed cameras, a timing unit, and a data storage and control computer. The primary and secondary racks are connected via a feed-through system which penetrates the external casing. The two racks are placed at an optimal distance from one another so that the vacuum line, fiber optic light guides, ethernet, TTL, and power cables all pass from one rack to the other.  Figure \ref{racks} includes a photograph of the racks, installed in their operational mode, inside the Zero-g aircraft. The lower panel of figure \ref{racks} provides a simplified illustration of the three experiments, oriented for simplest viewing angle, without cables, tubing, external casing, or the secondary rack. 

\subsection{Systems detail}

\subsubsection{Vacuum}
\begin{center}
\begin{figure}
\includegraphics[]{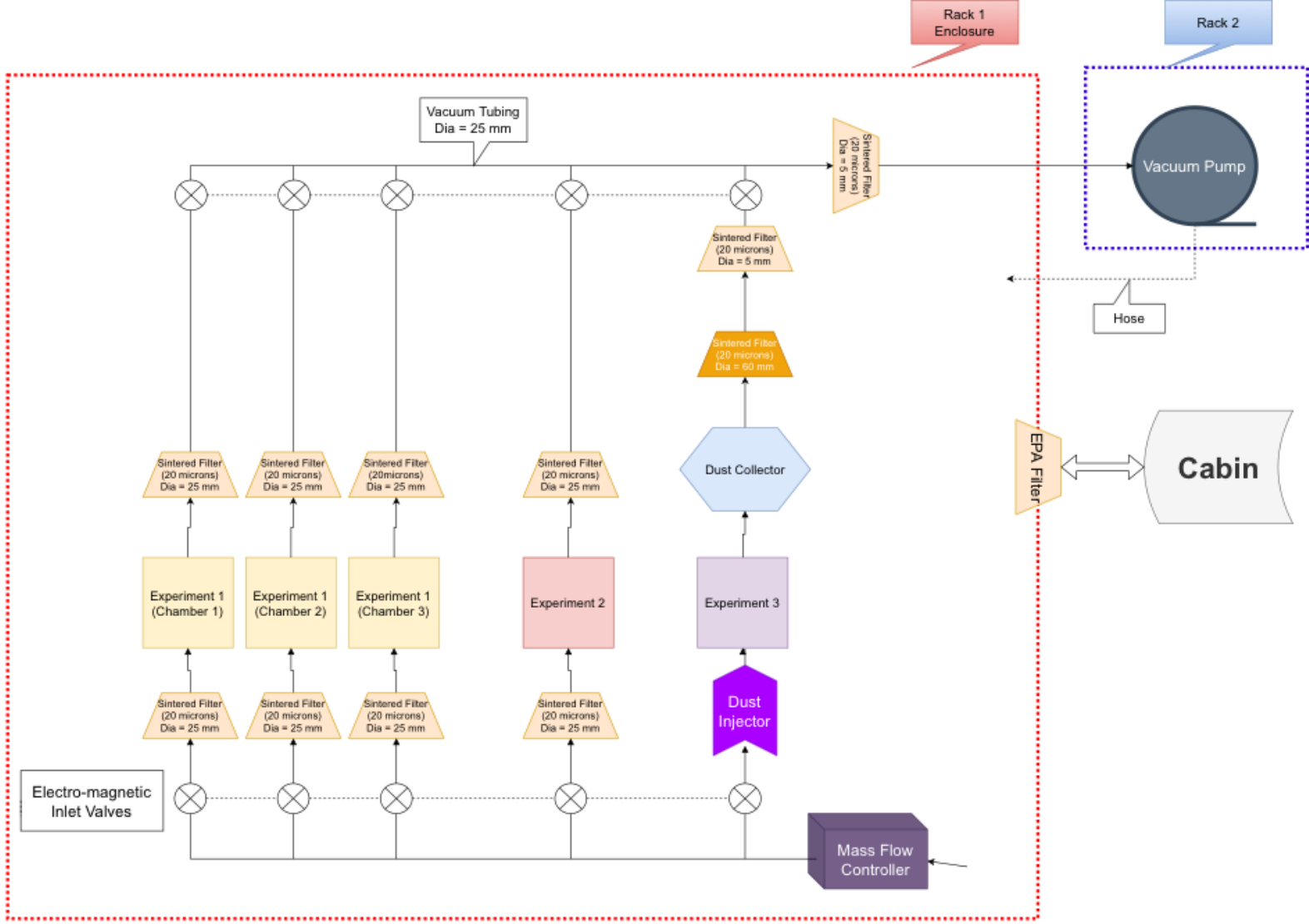}%
\caption{\label{vacuum_overview} Schematic view of the vacuum system in the TEMPus VoLA facility. We refer to the upstream direction as the flow originating from the Mass Flow Controller and the downstream direction as the location of the vacuum pump. The vacuum pump is held on a separate rack, and is connected to the primary structure through a vacuum tubing. The same vacuum line is shared by all experiments.  Each experimental chamber is enclosed by an upstream and downstream valve so that they may be selectively open to or closed off from the vacuum flow stream. The air from the aircraft cabin enters the containment unit through a high-quality filter. The air entering the mass flow controller is at ambient pressure, and is reduced at a prescribed flux rate, using the Bronkhorst Flowmaster control software suite. }%
\end{figure}
\end{center}

In figure \ref{vacuum_overview} we provide an overview schematic of the vacuum system. The secondary rack holds the vacuum pumping station, which connects to the primary rack by way of corrugated vacuum tubing. The entire system is evacuated by a (Pfeiffer HiCube 80 Eco, DN 63 ISO-K), consisting of a primary membrane pump and secondary turbo-molecular pump.
The flow is impeded at its entrance by a Bronkhorst mass flow controller (MFC), model F-201CV/F-211CV.  The MFC is placed upstream of all of the chambers and the communication to this device is sent to and received from the laptop computer via RS-232 in connection with a USB adaptor. 

A series of electrical valves separate the chambers so that the experiments can be effectively turned off: chambers with both up-stream and down-stream valves closed remain under vacuum, but when both the upstream and downstream valves associated with each chamber are closed, there is no net pressure gradient and thus no flow of gas through the chamber.  The electromagnetic valves are operated via a switch panel on the front of the primary rack. 

The pressure data logging system (Pfeiffer TPG 336) allows the simultaneous recording of six signals from Pirani pressure gauges, installed such that they can monitor the pressure in the individual chambers, at a frequency of 1 Hz. Since Experiment 1 involves a differential pressure measurement, we installed individual pressure gauges below each chamber (upstream of the granular media sample) and a single pressure gauge downstream of the three chambers.\ 

\onecolumngrid

\twocolumngrid  
\

\subsubsection{Particle tracking system}
The equipment in our particle tracking system consists of two high-speed cameras (Photron AX-100), a principle timing unit (PTU), a computer server, and two high-intensity pulsed LED light sources (LED Pulsing System PIV V3 from ILA5150). We control the equipment with the commerically available DaVis software from LaVision. Within this framework, we have the capability to apply more than one particle-based speed measurement technique. Indeed, for experiment 2, we require tracking of individual particles that are relatively large, and so we apply three-dimensional particle tracking velocimetry (PTV).  Because this experiment uses relatively large particles, we can apply backlighting to obtain images in shadow. The data processing requires an extrapolation of particle position from one frame to the next, in reference to a spatial mapping within the three-dimensional volume \citep{Bourgoin2006,Xu2008}. 

For experiment 3, we require small tracer particles and we are interested in collective patterns in velocity field, so we use instead particle image velocimetry (PIV). For particles below the pixel resolution scale, light scattering is required to detect the particles on the camera sensor. It is common to use a laser for this purpose, because the short pulses can deliver a lot of power in a short amount of time, enabling short integration times for the high-frame-rate recordings, with the result being minimal blurring of the particles in the images. However, lasers require extensive safety procedures and maintenance, and we find it more practical to replace the laser with two pulsed high-powered LED sources.  The cameras and light sources are synchronised via the PTU. We connect the light sources to the Q-switch line typically reserved for a laser. From this TTL signal, the light pulse frequency and duration follows the pre-programmed pattern.

In our setup, the light sources reside on the secondary rack, and we port the light pulses to the two experiments in the primary rack using fiber-optic light guides. For experiment 2, the light guides terminate in a panel screen which is used as backlighting for shadowgraphy. For experiment 3, the light guides reach a collimating light-sheet optic to generate a light-sheet for scattering measurements.  The feedthrough panel on the exterior of the primary rack gives us the option to switch the light guides to direct the light pulses to our chosen experiment. Diagrams showing how the light sources interface with the hardware are included in the descriptions of the experimental apparatus below.   

We require both fast (high-speed) measurements and relatively long duration covering the 20 seconds of the $0g$ phase of the parabola. Due to limitations in camera memory and speed of data write out-time, there is always a tradeoff between frame rate, recording time and resolution. Our system addresses this limitation with the capability for cyclic recordings, meaning that we can repeat high-speed recording intervals with a specified number of frames per cycle, and a specified repetition rate for the cycles. One therefore captures fast dynamical processes, but also has the option to space apart the measurements to cover systemic evolution over long duration.

\subsubsection{Experiment one: gas permeability chambers}

\begin{figure}
\includegraphics[width=0.5\textwidth]{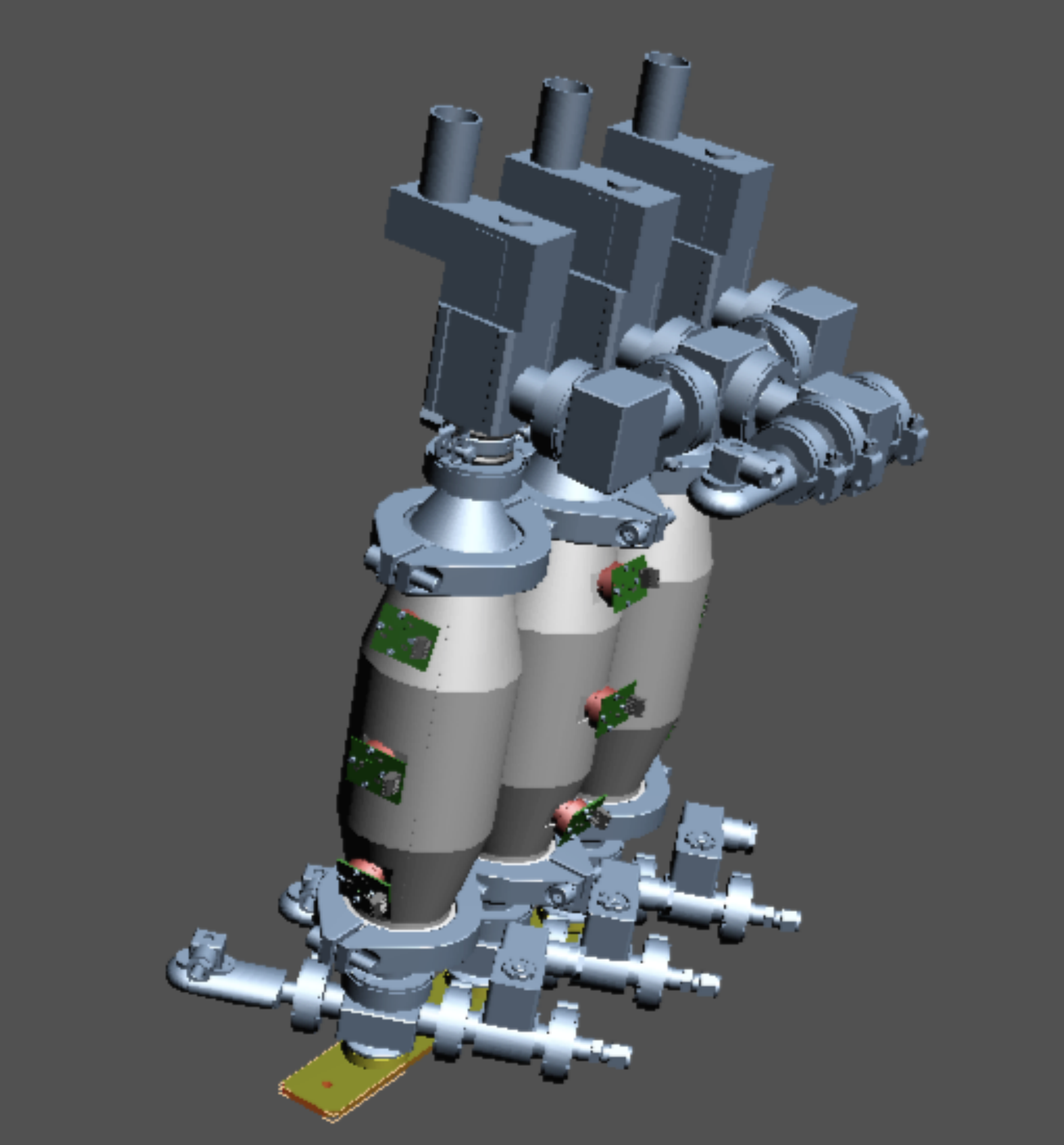}\\
\caption{\label{sys_1_cad} Isometric view of experiment 1, gas permeability of granular media under low gravity. Each chamber receives an individual gas inlet line (towards the right of the image) and the pressure is measured for each channel with separate Pirani pressure heads near the gas injection points. The three chambers are closed individually by six electromagnetic valves in total. The vacuum line at the tops of the chambers merges to a single t-junction, where a reference pressure transducer is installed. The cylinders are sealed by standard KF flanges, with sintered-metal centering rings used to hold the sample inside. The three containers are made of a transparent polycarbonate, so that the location and concentration of the sample can be monitored by both a camera and an array of photodiodes mounted on the sides of the containers.  }%
\end{figure}

The experimental vessel is a cylindrical canister packed with solid grains. 
The particles are trapped by vacuum centering rings containing sintered metal filters. The pore size of the filters can be easily changed depending upon the sample particle size. We have for example implemented 20$\mu$m mesh (Stainless steel 304/1.4301, DN 50 ISO-KF). The containers are made from a clear polycarbonate, to enable a view on the filling factor of the material inside. Figure \ref{sys_1_cad} shows the design of the system, including the gas inlet lines, the pirani pressure gauges, electromagnetic valves, and photodiode placement. 

The experiment is passive in the sense that it operates continuously through all phases of the parabola (the full range of gravitational loading, from $0 \le $g$_{load} \le$ 2g). We intend to study more than one particle type, and for this we add two additional canisters, so that the samples do not need to be exchanged during flight. We isolate the particle type under consideration via the valve system, which diverts the gas stream to the chosen vessel. 

\begin{figure}
\includegraphics[width=0.5\textwidth]{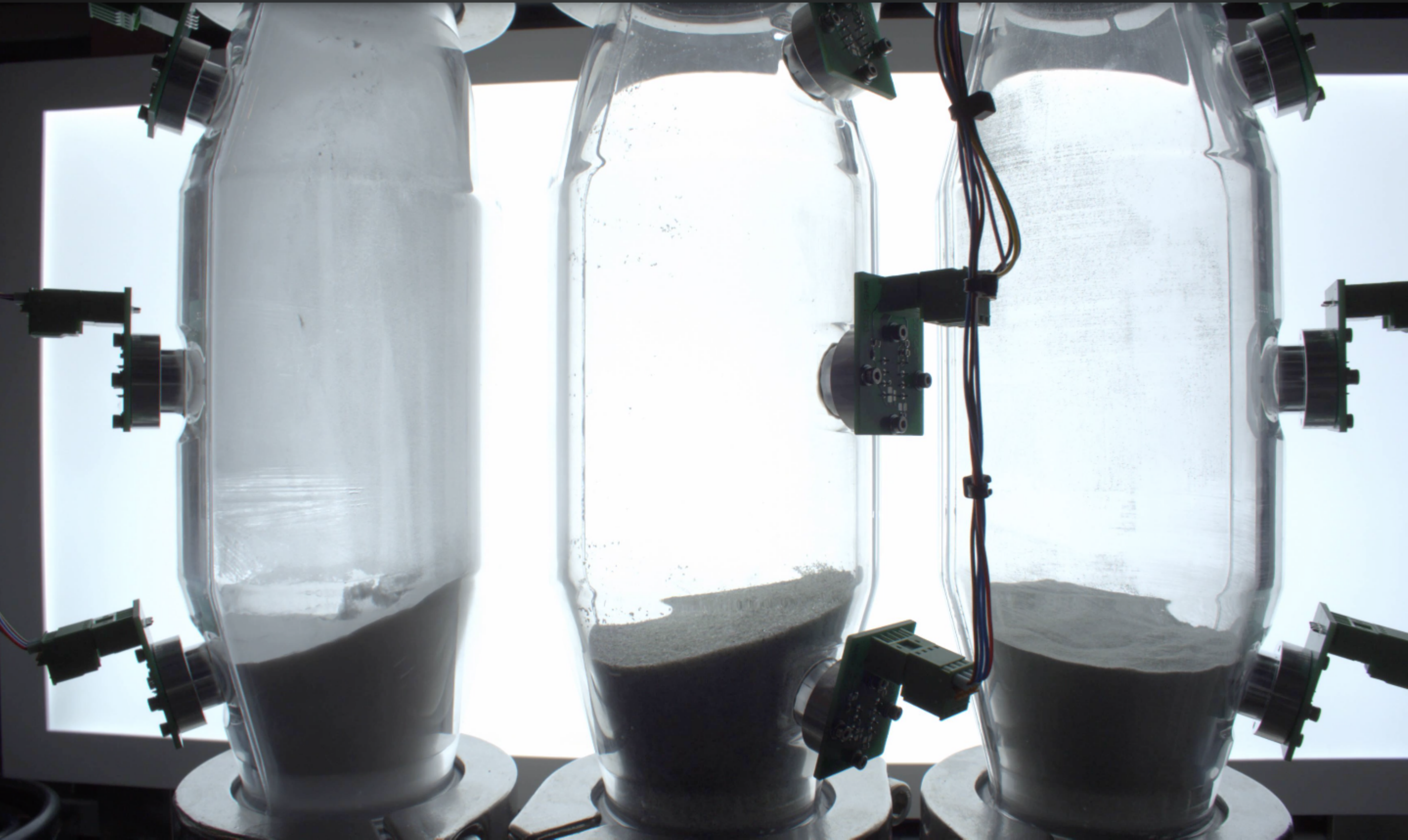}\\
\includegraphics[width=0.5\textwidth]{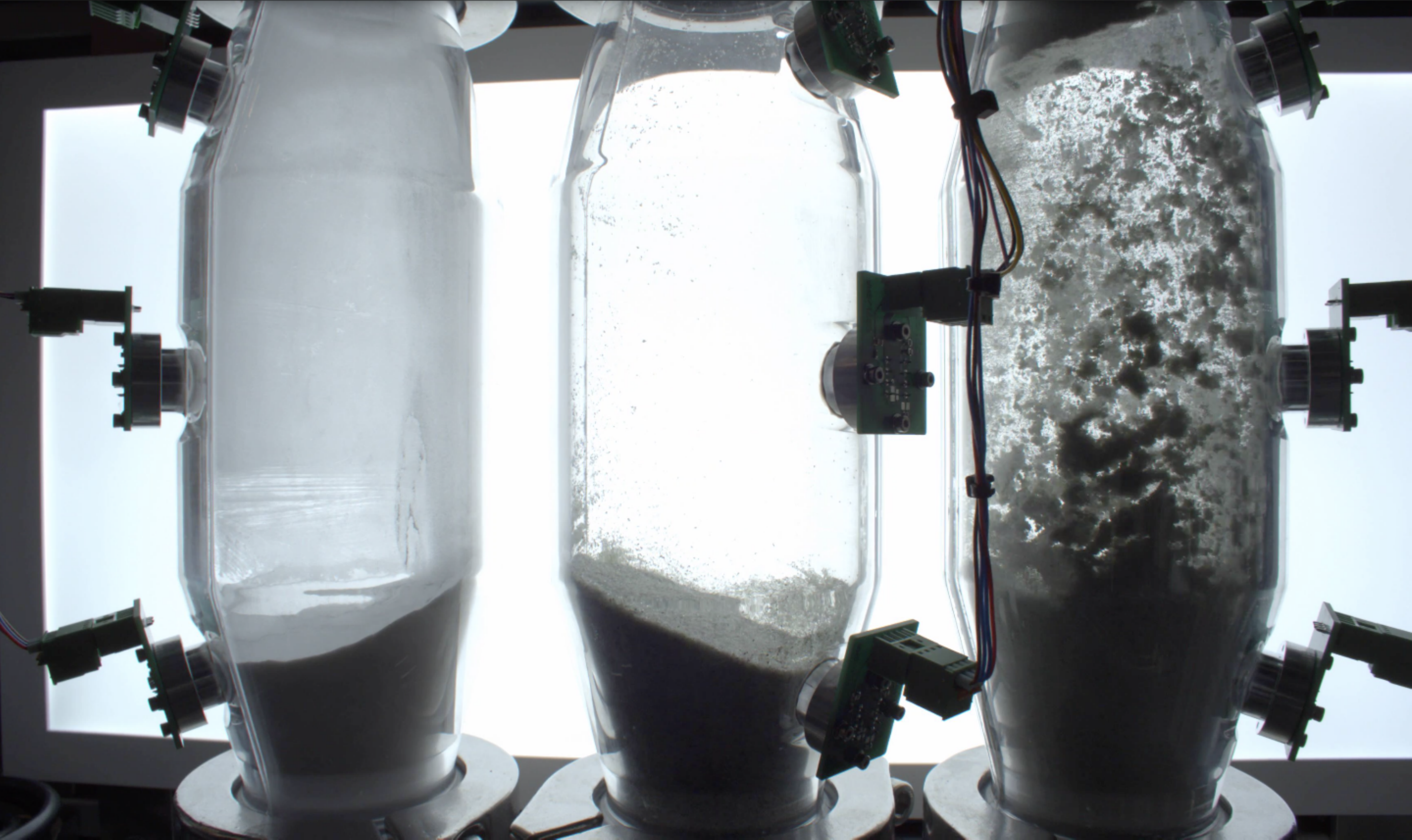}\\
\caption{\label{sys_1}The three chambers of experiment 1, during steady flight (top panel) and microgravity (bottom panel). The samples from left to right were: microgrit, olivine coarse ($\sim$ 500 $\mu$m), and olivine fine ($\sim$ 20 $\mu$m). The chamber on the right is the one in which the gas flow is active.}
\end{figure}

The quantities we measure are: differential pressure and sample optical depth (light transmission). 
Pressure is measured with Pirani pressure gauges.   We measure the transmission of light, originating from the LED light panel, in two ways: 1. with a high-resolution camera; 2. with an array of photo-diodes connected to a multi-meter/datalogger device.  Figure \ref{sys_1} shows an image obtained with the Thorcam camera. All three containers are shown, filled with particles of different sizes and compositions. In the images, the photodiodes mounted onto the external walls of the containers, can be seen. Note that they are placed at a position such that they receive light from the LED panel, but do not obstruct the viewing angle of the camera. There are 9 photodiode channels total, three for each chamber, placed bottom, middle, and top. The upper panel of Figure \ref{sys_1} is an image acquired during steady flight, corresponding to Earth's gravity. The lower panel of the figure is an image acquired during the zero-G phase of the parabola, while the third chamber was `active', meaning that the low-pressure gas was flowing only through this chamber. As expected, the particles fluidise and fill the chamber, thereby increasing the porosity of the sample.   

The imaging data is complementary to data from the photodiode array. The nine photodiodes are commercial products placed on circuit boards with built-in amplifiers (Texas Instruments OPT101 Monolithic Photodiode and Single-Supply Transimpedance Amplifier). The voltage is measured with a Keithley Multimeter-Multiplexor unit (models 2701-7710).  We operate the device in back-panel mode, using the Standard Commands for Programmable Instruments (SCPI) and Transmission Control Protocol (TCP), transmitted over ethernet cable.  We chose a time integration constant to minimise noise and maximise speed, resulting in a scanning rate of 27 Hz.

\begin{figure}
\includegraphics[width=0.5\textwidth]{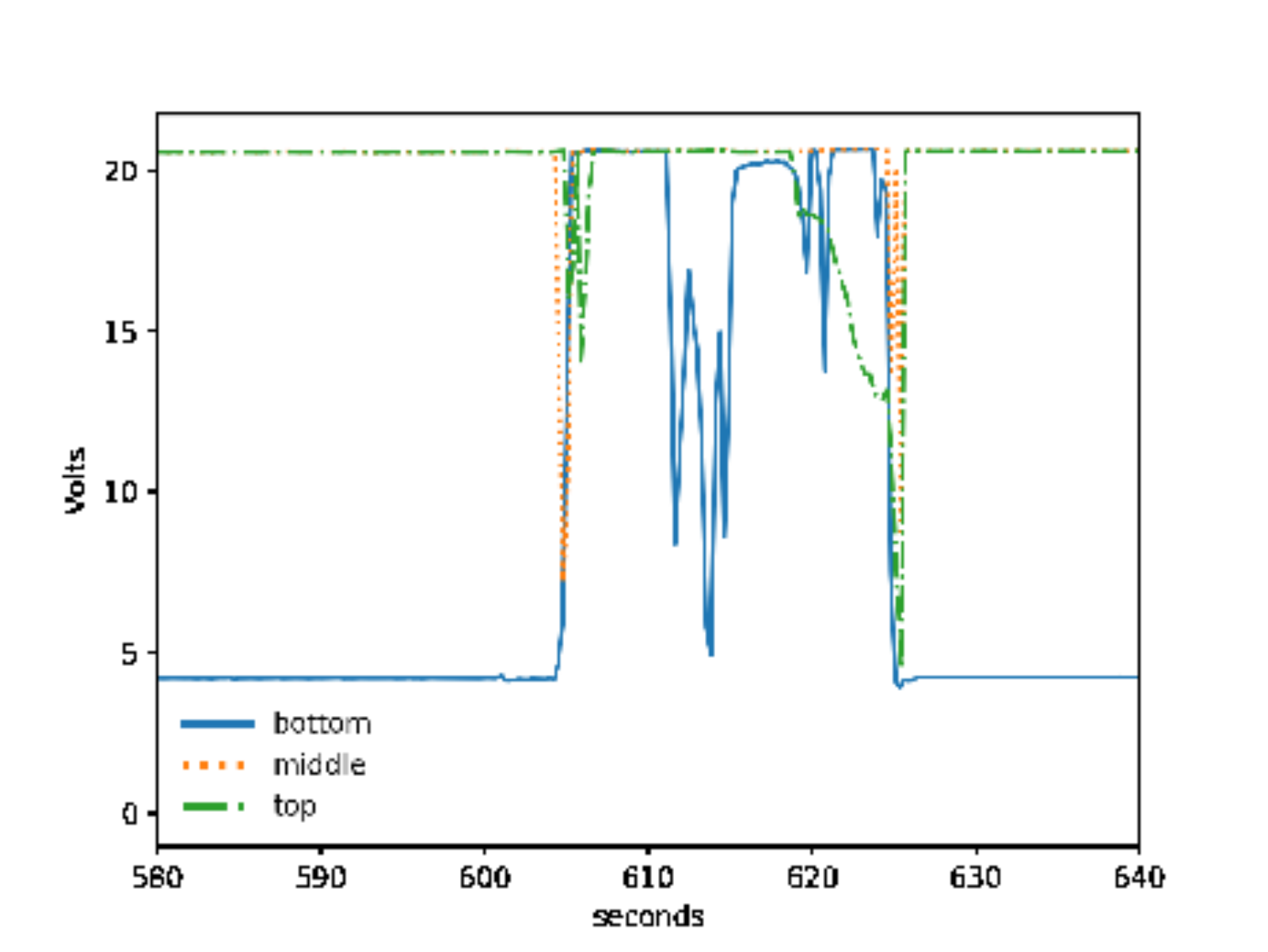}\\
\caption{\label{sys_1_multimeter} Example of a voltage measurement from the photodiode array attached to experiment 1, chamber 3. Blue solid line: bottom sensor; orange dotted line: middle sensor; green dot-dashed line: top sensor. As particles levitate and fill the chamber, the sensor on the bottom jumps to a value reflecting higher light transmission and therefore higher porosity. }%
\end{figure}

Figure \ref{sys_1_multimeter} demonstrates a typical signal from the photodiodes. We plot voltage vs. time for the three sensors attached to the third (right most in figure \ref{sys_1}) gas permeability chamber.  The duration over which variations in measured voltage occur match the time of a single parabola, lasting 22 seconds. It can be seen that the difference in voltage corresponds to expected changes in brightness: as the sample moves away from the bottom of the container, light is able to reach the lowest sensor and so the voltage increases. Note that the particles fluctuate their position due to jitter of the aircraft and so the signal is complicated. However, the time resolved multimeter data together with the movies from the Thorcam visible camera are sufficient to interpret the behaviour of the system and compare to the differential pressure measurements.  

Coincident with the dramatic change in light transmission in our experimental vessels, we also measure a dramatic change in differential pressure when the samples levitate. 
With our measurements of pressure, we demonstrate that:
\begin{itemize}
\item{the transient change in pressure when switching channels between chambers is quite short with respect to a single measurement over the duration of a given parabola;}
\item{the magnitude of the pressure differential is as predicted for a random packing of given particle size, at Earth's gravity;}
\item{the pressure differential remains stable while the bed is packed, and that fluctuations in $\delta P$ correspond to variations in gravity. }

\end{itemize}

To explore variations in $\delta$P as a function of gravitational loading, we must first synchronise the signals from the pressure transducers and the accelerometer. We utilise the acceleration data measured from the aircraft cockpit and the logged pressure from our vacuum equipment. The timestamp from the two devices is synchronous only on the minute precision level. We check the exact offset by performing a cross-correlation between the pressure and accelerometer signals. This is a standard procedure in time series analysis, by which one iteratively applies a small temporal shift to the time series, and calculates the correlation with a corresponding signal, for each shift in time, $\tau$. The result is a correlation curve as a function of $\tau$. The maximum of the correlation function corresponds to the best guess for the offset in time. For example, the correlation peak calculated in this fashion turned out that the delay time with the highest correlation is $\tau=10$ sec.  

\begin{figure}[ht]
\includegraphics[width=0.5 \textwidth]{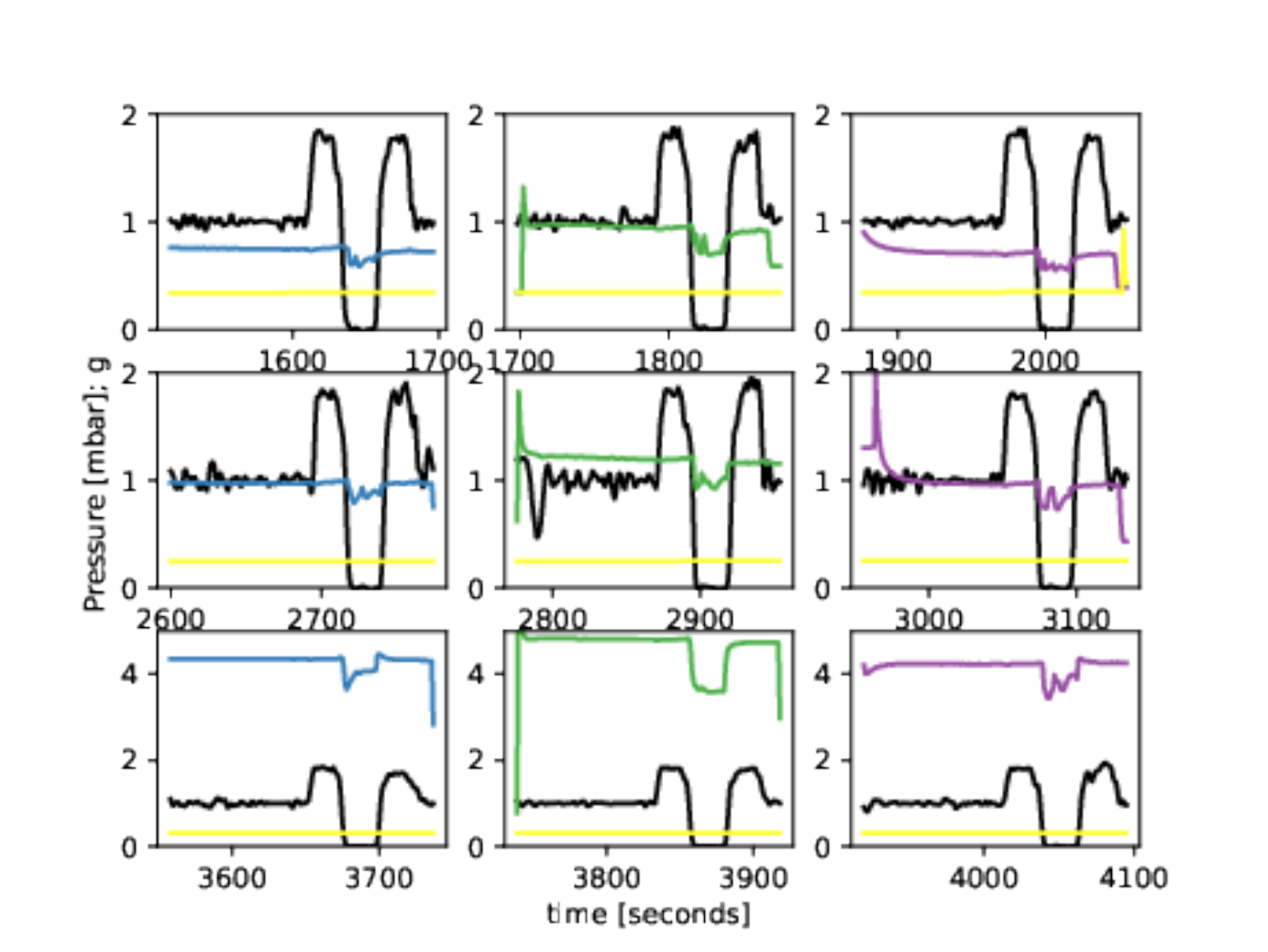}
\caption{\label{pressure} From experiment 1, measurements of gravitational acceleration and differential gas pressure for 9 parabolas. The yellow line is always the pressure measured downstream of the sample. First column: Experiment 1 chamber 1, sample material is glass beads. Blue curve is pressure upstream of the sample. Second Column: Experiment 1 chamber 2,  sample material is olivine sand.  Green curve is pressure upstream of the sample. Experiment 1, chamber 3, sample material is steel spheres. Purple curve is pressure upstream of the sample. Top row: mass flux rate is 0.01\%; Middle row: mass flux rate is 0.05\%; Bottom row: mass flux rate is 8\%. The magnitude of differential pressure increases with increasing gas mass flux at all levels of $g$. Fluctuations in the pressure coincide with the $0g$ phase and decrease since the particle porosity decreases when the gravitational force is absent. }
\end{figure}

Figure \ref{pressure} demonstrates the validity of our method to synchronise the acceleration and pressure data. This is an example of a measurement sequence over 9 parabolas, where columns 1, 2, and 3 correspond to experiments conducted in permeability chambers 1, 2, and 3, respectively. The rows in this figure, upper, middle, and bottom correspond to differences in the prescribed mass flux rate, from low to high (always less than 10 percent of the full range).  We plot the gravitational acceleration in units of Earth's gravity $g$, where the changes from steady flight to hypergravity (up to 2g during pull up and pullout) and to $0g$ are all clearly discernible. We overlay the measurements of the pressure, where the lower (yellow) curve is always the pressure measured above (downstream of) the sample. The coloured curves are the pressure measured above the samples, all with particle sizes $\sim$500$\mu$m. In this case, blue corresponds to glass beads, green corresponds to coarse olivine grains, and purple corresponds to steal spheres. As expected, the pressure differential shown in Figure \ref{pressure} is greater when the mass flux increases. Moreover, the total duration of the fluctuations in pressure matches the duration of the Zero-g phase.  The pressure variations for these and additional samples will be the object of an ulterior publication.

\subsubsection{Experiment two: dust-drag chamber}

\begin{figure}[ht]
\includegraphics[scale=0.3]{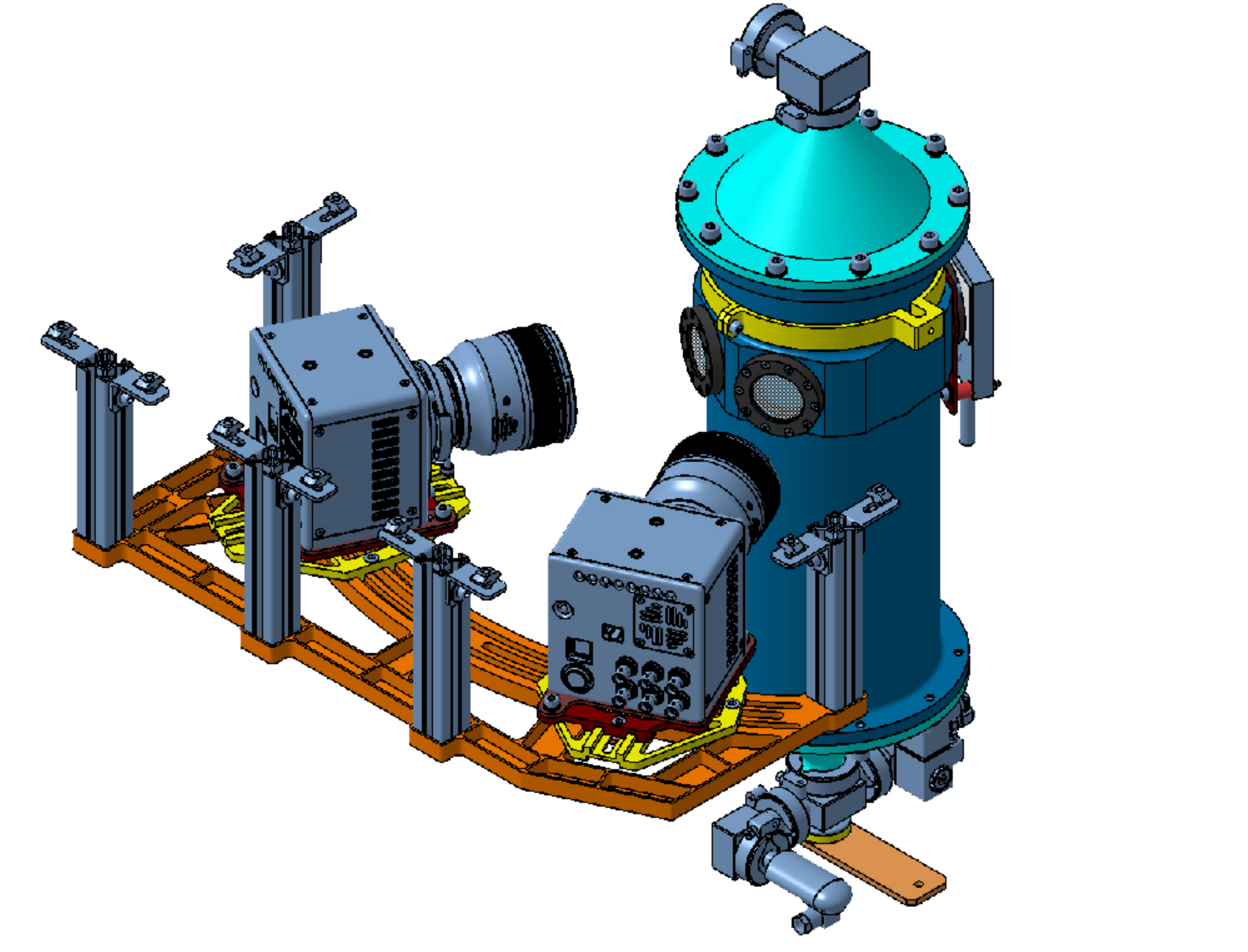}%
\caption{\label{sys2_diagram} Rendering of experiment two apparatus and measurement system. The chamber is dedicated to determining aerodynamic drag forces on particles. Cameras are placed at a 30 degree angle from one another and are both focused on the centre of the chamber. They are placed on a rail to enable fine adjustments to their position.  A panel light is mounted directly opposite each camera to provide back-lighting. Pressure is measured at the gas inlet. The top of the chamber connects to the vacuum line.}%
\end{figure}

This experiment is oriented vertically with respect to Earth's gravity, and involves an empty chamber of 10 cm diameter. Figure \ref{sys2_diagram} illustrates the setup.  
Like experiment 1, there are sintered mesh centering rings on both ends of the chamber to trap particles inside. The purpose of the chamber is different, however, since we are interested in measuring the trajectories of individual particles, and we do not want the particles to disturb one another. So the number of particles comprising the sample is relatively low: a few 10s of particles pre-loaded into the chamber. We load the particles by removing the top flange and pouring the particles into the open chamber, before sealing it again. The gas pressure is measured near the bottom of the chamber, which is the side where the gas stream enters.  The vessel is evacuated from the top. Four viewing windows are built into the side of the chamber, the two cameras are placed such that they provide two viewing angles on the centre of the chamber, to enable a stereoscopic (three-dimensional) recording of the particles when they levitate. Panel lights are mounted externally and directly opposite to the cameras, to provide backlighting. The light in the panels originates from the pulsed LEDs on the secondary rack, and arrives by way of a fiberoptic light guide. 

\begin{figure}[ht]
\includegraphics[scale=0.29]{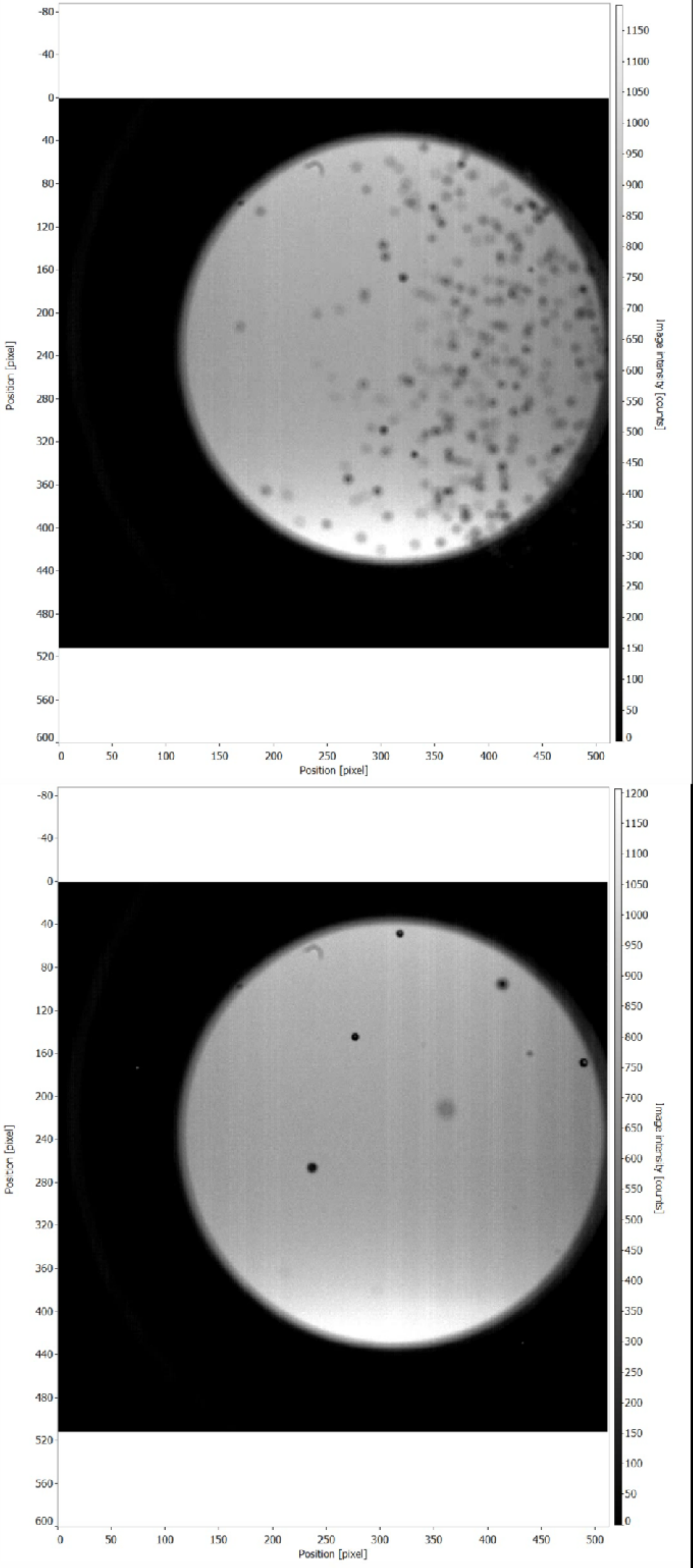}

\caption{\label{drag_levitation} From experiment two, examples of particles imaged with high-speed cameras using shadowgraphy. Both axis are pixel position and the greyscale image intensity is displayed in counts. The top image is an example of an instance where many particles are in the field of view. The bottom is an example of very few particles. The images demonstrate that we can capture individual particle dynamics even with very little sample material, as in the top panel. It is also possible to also study relatively dense particle seeding populations as shown in the bottom panel, which is however of less relevance for protoplanetary discs.}%
\end{figure}

In figure \ref{drag_levitation}, we show two examples of the seeding density of particles that were measured in the view of one of the cameras during the microgravity phase. The top panel shows a shadowgraph image of particles when the seeding density is at its maximum. The bottom shows lower seeding density, recorded at a later time. In either image, we are able to identify individual particles. We find however, that the acceleration drift of the aircraft affects our ability to detect the same particles simultaneously in both windows. We addressed this matter with a modification to the position of the cameras, such that the axis bisecting both the chamber and the cameras is parallel to the direction of greatest acceleration drift. 

This has been a demonstration of the principle using simplified test particles, namely, glass spheres of 0.5 mm. Drifting pebbles that participate in fluid instabilities in protoplanetary discs necessarily have a high relative particle-gas velocity and their drag coefficient depends upon Knudsen number. The pebbles in a protoplanetary discs are not solid spheres, but rather porous aggregates and so gas can flow not only around but through them. The slip drag coefficients for porous aggregates should depend on both Knudsen number and the permeability coefficient of the porous aggregate. We give a full literature review and new analysis of dust aggregate morphological types in the companion work by Bodenan et al. 2022 (in submission). In future campaigns, we will advance to studying particles of more realistic dust aggregate morphologies. 

\subsubsection{Experiment three: two-phase shear flow chamber}

\begin{figure}[ht]
\includegraphics[angle=270,width=0.5 \textwidth]{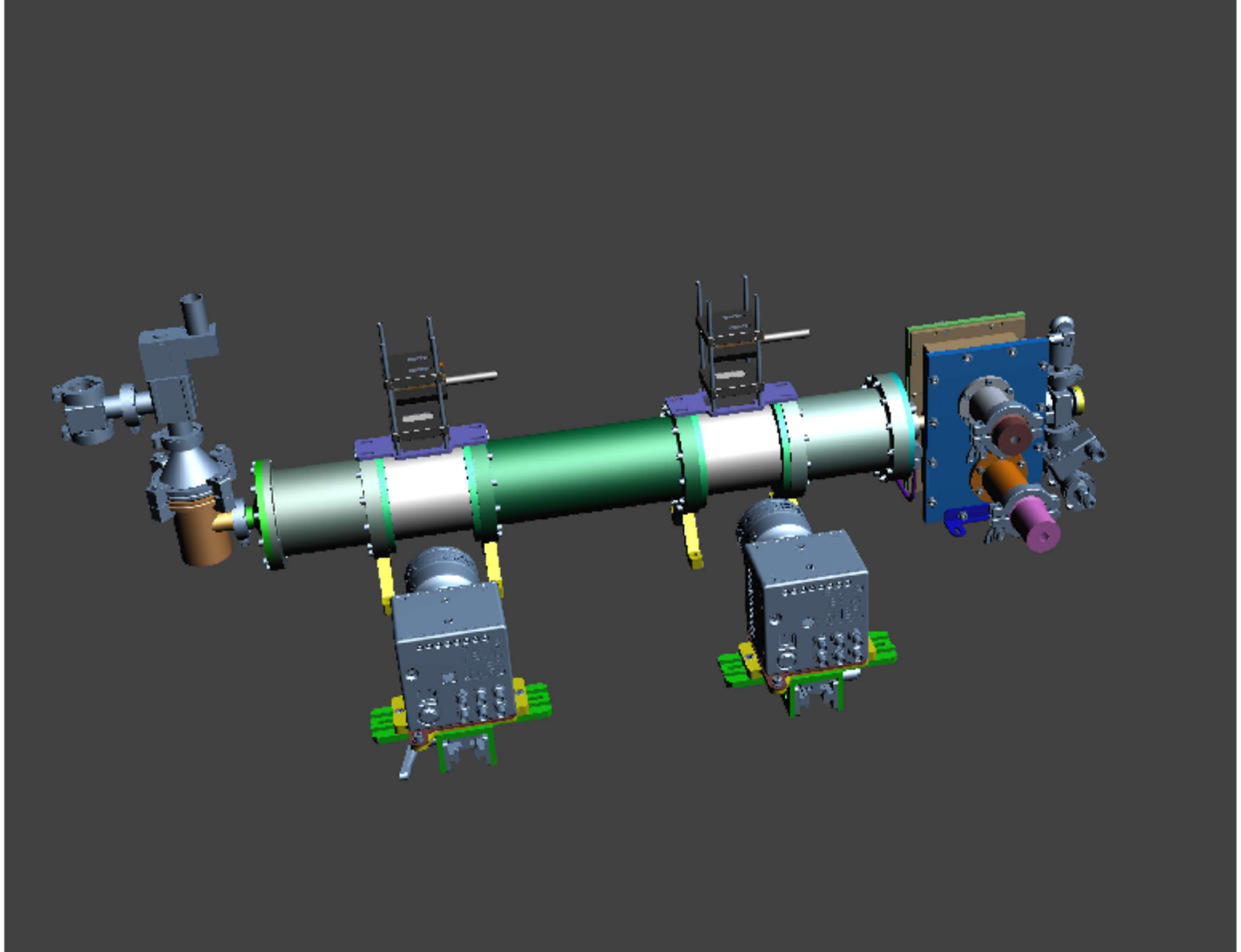}
\caption{\label{sys3_tube} Rendering of experiment three, dusty shear flow chamber. To the right of the image is the dust injector  (detail in Figure \ref{dustinjector}) and gas inlet line. To the left of the image is the dust collector. High speed cameras are placed next to two transparent segments at an upstream and downstream location. The optics mounted on top of the transparent segments create light sheets that are parallel to the camera sensors (detail in Figure \ref{optical_illustration}) }
\end{figure}

\begin{figure}[ht]
\includegraphics[width=0.47\textwidth]{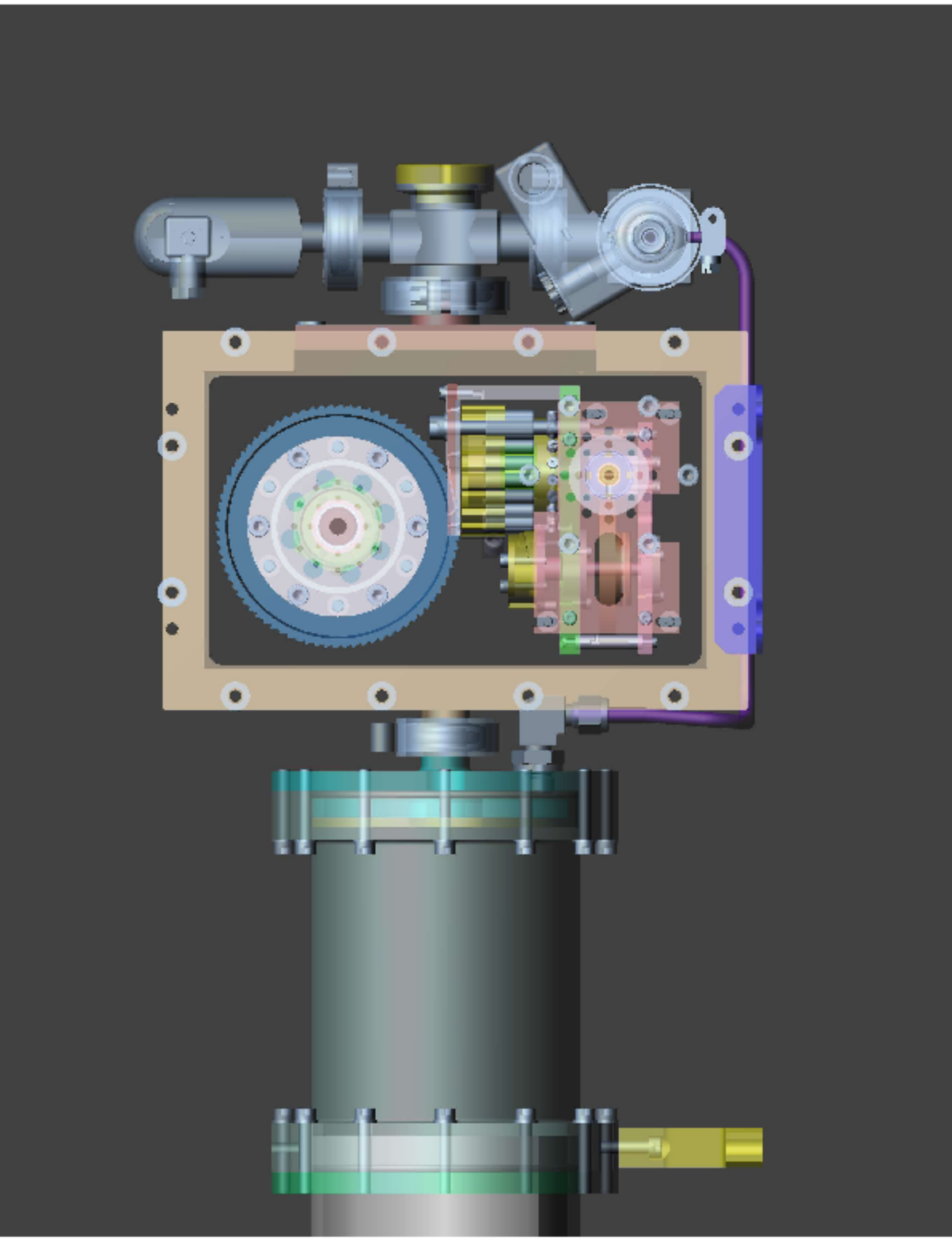}
\caption{\label{dustinjector} Closeup view of the dust injection mechanism used in experiment three, to seed the mid-line of the vacuum shear flow with dust particles.  The large circle is the dust dispersal wheel, operated by one motor, the cog-wheel and revolver unit in pink, green and yellow is where the dust holds the preloaded dust until it is introduced by a levitating piston.}
\end{figure}

\begin{figure}[ht]
\includegraphics[width=0.5\textwidth]{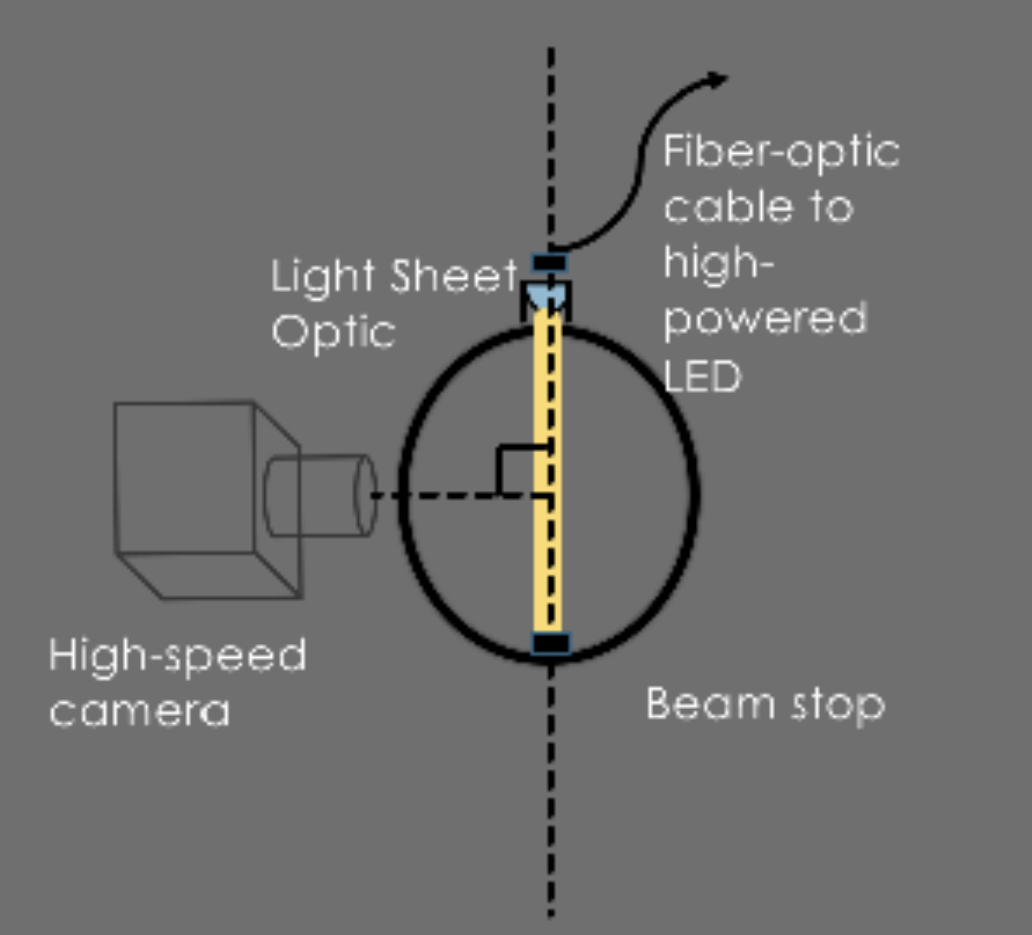}

\includegraphics[angle=270,width=0.5 \textwidth]{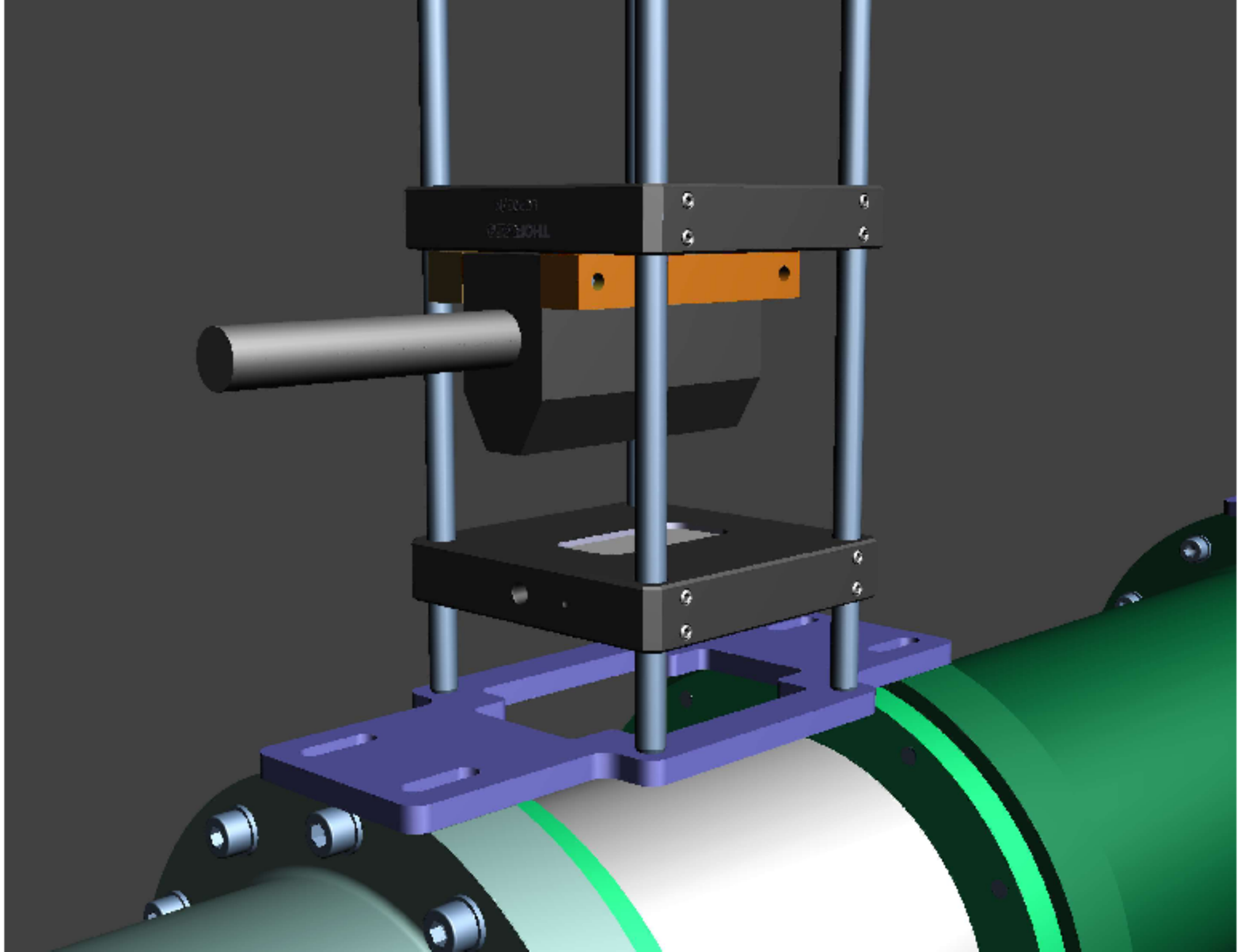}

\caption{\label{optical_illustration} Top: cross-sectional schematic showing the orientation of the light sheets and cameras in experiment three. The light sheet illuminates a slice through the middle of the flow vessel, creating a two-dimensional plane for studying the flow dynamics. Bottom: The light sheet optics mounted above the clear segment in the flow vessel. The rod represents the fiber-optic cable through which light enters. The light is then collimated through a slit, and focused through a cylindrical lens.}
\end{figure}

\begin{figure}[ht]
\includegraphics[trim={3cm 1.9cm 3.75cm 0},clip, width=0.45\textwidth]{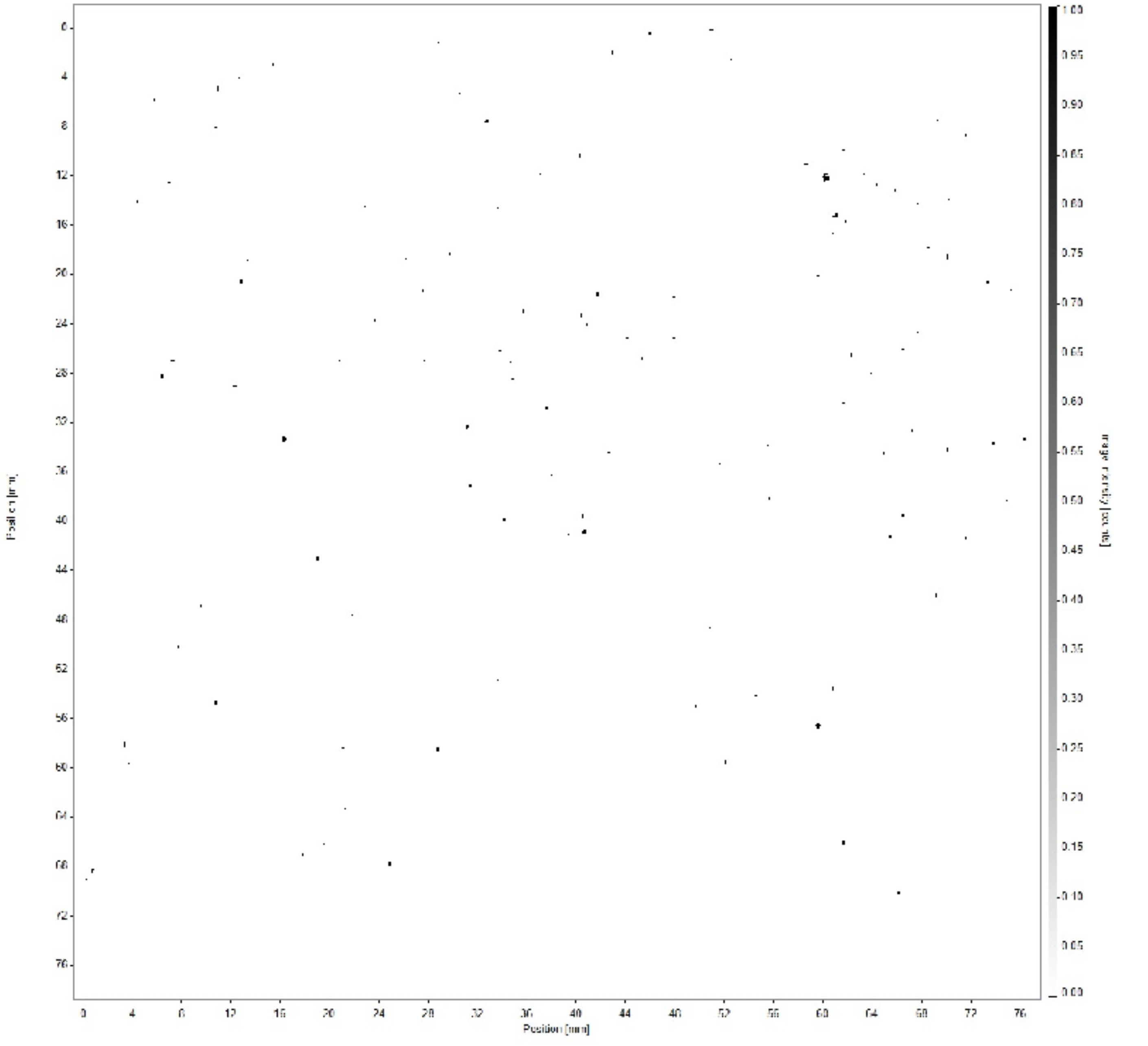}
\caption{\label{raw_lightscattering} Example of the raw data product from experiment three. The single image from the high-speed movie sequence shows light scattered off of dust particles, collected during the microgravity phase of a parabola. The field of view is 67$\times$67 mm$^{2}$. The image intensity values are inverted for viewing purposes. The image greyscale ranges from 0-1. The particles can be clearly identified despite illuminating no more than a few pixels. The faintness of the particles is an indication that the dust particles are in general disperse and that agglomeration due to electrostatic forces is not the prevalent phenomenon.}
\end{figure}

This chamber contains a gas flow with target velocity of ~1 ms$^{-1}$ and pressure of 0.1-10 mbar. The center-line of the flow is seeded with dust particles of sizes 1-15 $\mu$m, originating from a dust injector. 

The chamber, upstream dust injector, and downstream dust collector, can be seen in Figure \ref{sys3_tube}.  Also pictured are two clear segments in the tube which enable high-speed recordings of the particle motions. The two video cameras operate simultaneously with the light sources mounted above the segments.  

A closeup view of the dust injection mechanism can be seen in Figure \ref{dustinjector}. The top panel illustrates two revolving mechanisms: the first is a set of twenty chambers that are preloaded with dust, which is gradually pushed out by a cog-wheel and piston; the second is a wheel that deagglomerates the dust and distributes it into a cross-flow. The dust travels into the centre of the chamber through a flange. 
The dust injector is operated by two Faulhaber motors. 

The maximum voltage that can be applied to the motors is 12V. We find that the maximum rotation frequency of the dispersal wheel is 100 Hz. The lower mechanism which lifts the powder in front of the dispersal wheel rotates much slower than 1 Hz. We find that the time it takes to empty one chamber and switch to the next chamber is 1min 45s. As there are 20 chambers, we can therefore create 35 minutes of continuous dust stream. The delay when changing between chambers can last a few seconds, but is shorter than a micro-g interval.  In order to time the release of the dust during the $0g$ phase of the parabola, we use a switch on the front panel of the rack to interrupt the power to the motors when the experiment is not operating. 

We expect for the dusty flow to develop a shear profile and we monitor its evolution. Moreover, we expect for the flow to be axis-symmetric and therefore we can obtain a clear understanding of the particle-gas dynamics by simply studying a cross-section of the tube. We achieve this by generating a light sheet that bisects the tube and is parallel to the plane of the camera sensor. Figure \ref{optical_illustration} shows the principle of the camera-illumination system. The schematic illustrates that we effectively study only a slice of the chamber by illuminating a volume that is nearly a plane. The light-sheet generating optics are also apparent in the figure. The source of photons is the same set of pulsed high-powered LEDs which illuminate experiment 2, and which reside on the secondary rack. The operator is able to switch the lights between system 2 and 3, by moving the light guides that join the two racks. We find the light-sheet beam thickness to be 4mm and for the divergence angle to be negligible within the distance that we illuminate.

We present an example of the data product that results in figure \ref{raw_lightscattering}. The image is colour-inverted for viewing purposes. 
The light scattered off of the $\sim10\mu$m particles occupies only a few pixels. We have verified that we can perform particle image velocimetry (PIV) with the data products at the given frame rate of 1KHz. We used the commercially available DaViS 10.2.0.74211 software to perform the image processing. The processing involved interrogating image pairs with a 2-pass PIV scheme, with both passes using $64 \time64$ pixel windows, to ensure sufficient particle density, with a 50\% overlap.

We show an example of the flow field resulting from PIV in the next section. We present a thorough analysis of the flow vector field, in comparison with predictions from numerical simulations, in a separate publication (Capelo et al 2022B, in prep).

\section{Verification of flow conditions}\label{verification}

We specify the dimensions of the experimental vessels such that the gas pressure and flow speed will match specific values. The primary calculation is with regard to mass flow conservation, taking as input variables:  the vacuum pump throughput T, the mass flux Q which is controlled by the Bronkhorst MFC, and the cross-sectional area, a, of each flow vessel. The throughput is defined by the performance of the vacuum pump and is invariant to pressure. T is given by the quantity  T$=$av, in dimensions of cubic volume per unit time, and v being the flow speed. For example, the fixed mass flow rate Q is expressed via conservation as Q =$\rho$av=$\rho$T.  In the product, $\rho$ denotes gas density, a the cross-sectional area of the pipe through which the gas flows. We aim for flow speeds in the range $1 <$ v $ < 10$ ms$^{-1}$. The desired gas pressure, related to the density by the ideal gas law, is then achieved by the prescribed combination of Q and flow vessel diameter.

\subsection{Experiment 1: Gas permeability}
We would like to check how much powder, of a given porosity, is required to induce a measurable pressure difference when gas flows through it. We make initial estimates based upon simple Navier-Stokes equation with linear drag force, that is mediated by a permeability coefficient $\kappa$. We impose conservation of mass flow rate, Q. The subscript u refers to `upper' pressure and l to `lower' pressure, above or below the granular bed. 
\begin{equation}\label{continuity}
\rho_{u}a_{u}v_{u}=\rho_{l}a_{l}v_{l}=Q\\
\end{equation}
\begin{equation}
\rho_{u}a_{u}v_{u}=\rho(h)a(h)v(h)\\
\end{equation}
\begin{equation}
\frac{dP}{dh}=-\frac{\mu}{\kappa}v(h)\\
\end{equation}
Via the ideal gas law, P can substitute for $\rho$ in equation \ref{continuity}, which gives
\begin{equation}
v(h) =\frac{v_{l}P_{l}a_{l}}{P(h) a(h)}\\
\end{equation}
\begin{equation}
\int_{P_{u}}^{P_{l}}{P(h)}dP=-\frac{\mu}{\kappa} v_{l}P_{l}\int_{h(P=P_{u})}^{h(P=P_{l})}{}dh.\\
\end{equation}
After integration, we get:

\begin{equation}
\kappa= \frac{\mu\delta z}{|\delta P | \frac{P_{u}+P_{l}}{2}} \frac{Q}{a_{l}}\frac{RT}{M}
\end{equation}

Here R, T, and M are the gas constant of value 8.314 m$^3$ Pa K$^{-1}$ mol$^{-1}$, temperature, and mean molecular mass, respectively. $\delta z$ is the height of the bed. By measuring P$_{l}$ and P$_{u}$, and with known sample height and mass flow rate (or linear flow rate, v), $\kappa$ can be extracted. The units of this quantity are m$^{2}$ and it refers to the average cross-sectional pore opening space (i.e. the dimension of the channels through which the flow can pass). We do not know the value of the permeability a priori, but for spherical particles and a flow with low Reynold's number, the equivalent quantity can be derived using the Carman-Kozenzy correlation, $ \kappa=K_{coz} d_{p}^{2}(1-c)^{3}/c^{2}$, where c is the porosity, and related to f, the filling factor, $c+f=1$. For a random-packed bed of spheres of equal size, f is always 0.64 and $K_{coz}$ is 2/90.

We test the pressure drop we should expect with height, which is described by the equation: 

\begin{equation}
P(h)=\sqrt{P_{l}^{2}-\left(\frac{2\mu v_{l}P_{l}}{\kappa}\right)h}.
\end{equation}

\begin{figure}[ht]
    \centering  
        \includegraphics[width=0.5 \textwidth]{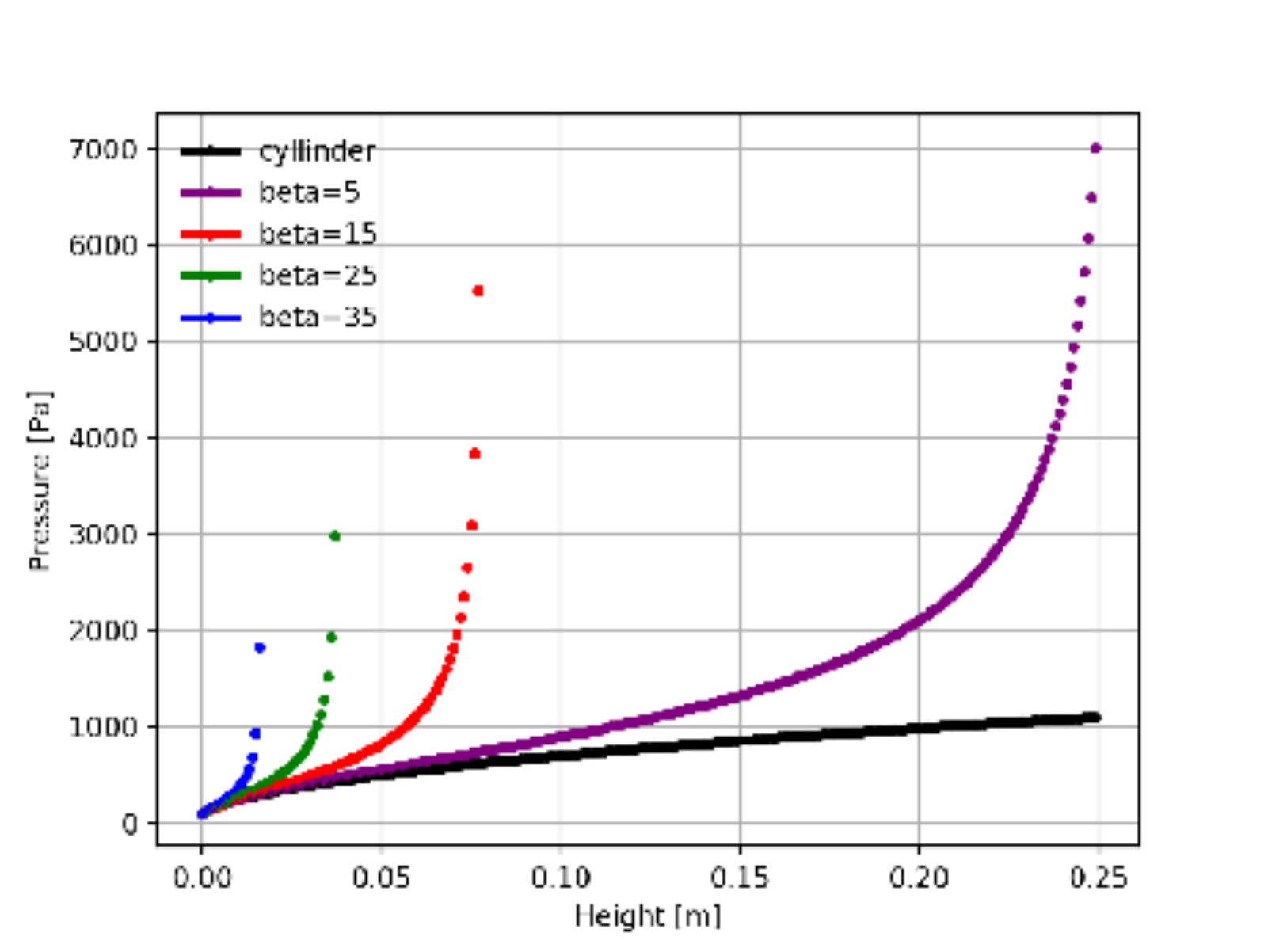}\\
        \includegraphics[width =0.5 \textwidth]{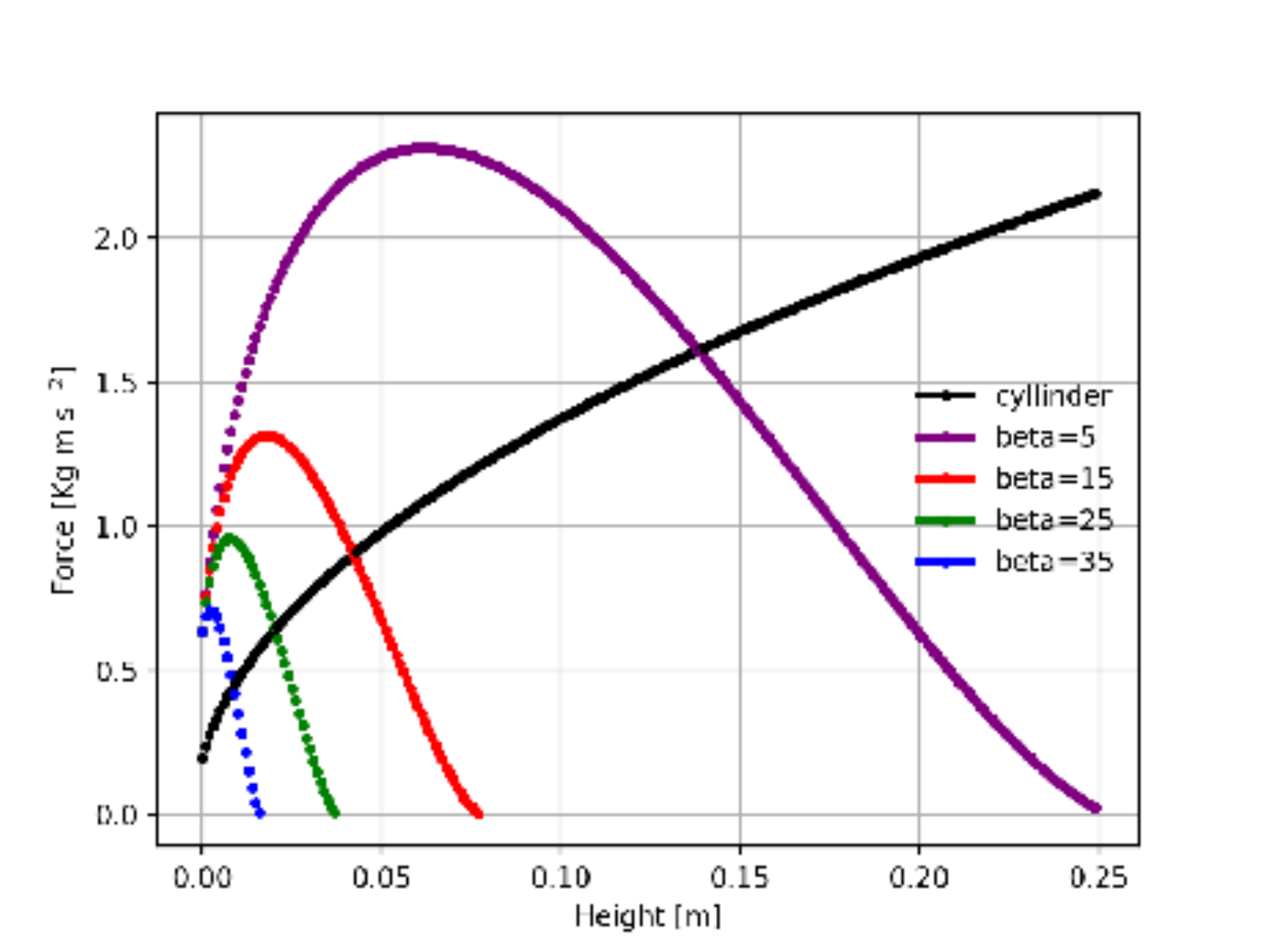}
    \caption{\label{forceprofile}Top: pressure profile in the packed granular bed, assuming random packing. Bottom: force profile in the sample for different conical opening angles, $\beta$. A value of $\beta=15$ was chosen so that an easily measurable pressure drop (approximately 8 mbar) would occur for a sample height of $dz=5$ cm. The corresponding force profile shows that for a conical shaped vessel, the sample will experience maximum force  near the top to middle of the sample, whereas for a cylindrical shape, the sample will always have maximum force at the top. See especially that the red curve corresponding to $\beta$=15 peaks very close to a force equal to Lunar gravity, representing the threshold where the pressure differential force and the gravity force are in balance. }
\end{figure}

It is common to assume a correction to the permeability of the form: 
\begin{equation}
\kappa_{\rm Klinkenberg}=\kappa\left(1+\frac{b}{\frac{P_{u}-P_{l}}{2}} \right). 
\end{equation}
The exact form of  the correction term $b$ is highly debated \citep{LASSEUX2017660}{}, but an example of the relation ship is: $b=0.7\kappa ^{-0.3}$. We see that the Klinkenberg correction becomes negligible if the order of the permittivity is very small compared to the order of the pressure differential. In Figure \ref{forceprofile}, we consider a case where the black line corresponds to $\kappa=7.5\times10^{-10}$, and the mean pressure is  of the order 500 pa. Therefore, we consider it justified to ignore the correction term and verify the performance of our apparatus using only the Karman-Cozeny formulation, for the set of data corresponding to the relatively high mass flux rate and large pressure differential; that is, the continuum assumption is nearly valid.

As we can predict the pressure profile within a sample, we can also predict the gradient in force with changing height. Figure \ref{forceprofile} shows a calculation of the force on the sample due to the flow. When the gravitational acceleration is less than this force, we expect the sample to break apart and fluidise.

We calculate the fluidization threshold of particles by assuming that the two forces balancing one another are gas drag and gravity, and that the force on each layer in the bed, can be given by multiplying the pressure by the cross-sectional area. When the force is in excess of the gravitational acceleration, the layer should lift, thereby increasing the porosity. 

Figure \ref{forceprofile} shows the force on the sample as a function of height, assuming the pressure profile depicted in the upper panel of the figure, and a granular bed of height 5 cm. Since the flow moves from high to low pressure against gravity, the highest pressure, and highest force, is at the bottom of the sample when a cylindrical geometry describes the cross-sectional area. One would then start to push the sample up from the bottom, when the gravity reaches 10-20 \% of g.

We re-considered the geometry to invert the forcing profile. We therefore revisit Eq. \ref{continuity}, and assume that a(h) is not constant, as in a cylinder, but changes, as in a cone, so that the area varies as a(dz)= $\pi(r-dz tan(\beta))^2 $. We recalculate the pressure drop and see that it changes the pressure profile in the sample dramatically, and hence also inverts the force profile.  
Considering the curve corresponding to a conical shape with opening angle $\beta=15$, we see that the force on the sample (about 6 cm) is not maximal at the bottom, but rather closer to the top of the sample. 

During the entry into microgravity, the transition is too rapid to watch the sample breakup. However, we present data from a partial gravity experiment during which the pilots maneuvered the aircraft to achieve lunar gravity. This value of 1/10 Earth's gravity was well matched to the fluidization threshold of our samples as shown in figure \ref{forceprofile}.
\begin{figure}[ht]
    \centering  
        \includegraphics[width = 3.5cm, height = 7cm]{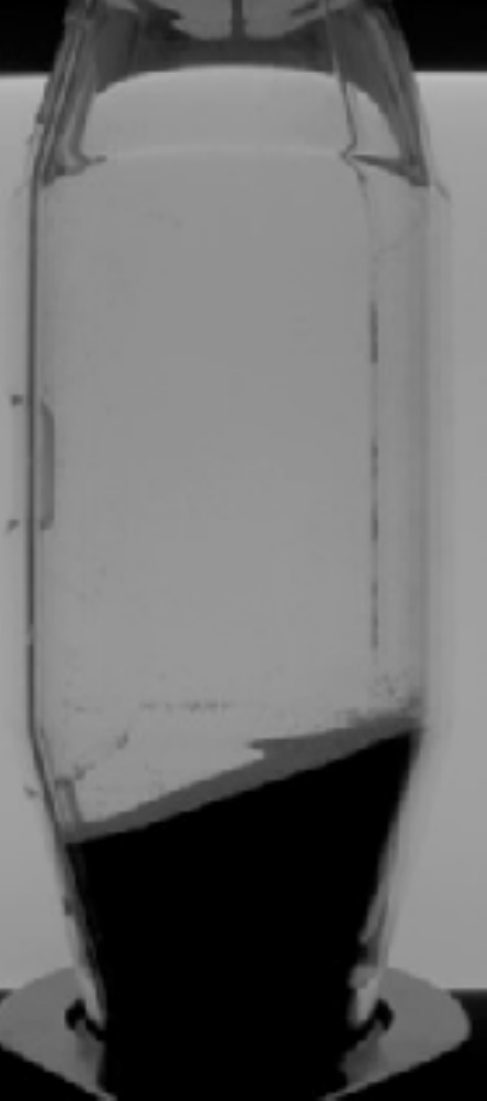}
        \includegraphics[width = 3.5cm, height = 7cm]{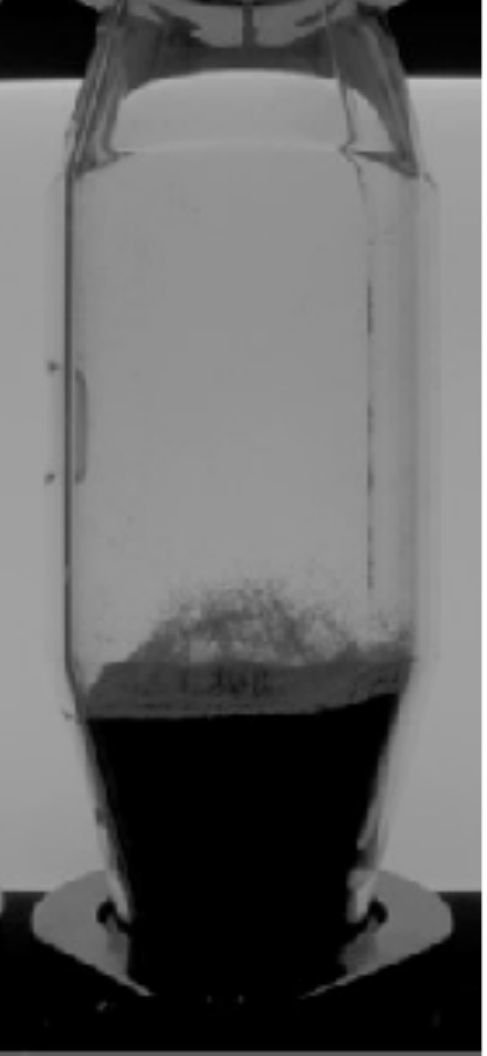}
        \includegraphics[width = 4cm, height = 4cm]{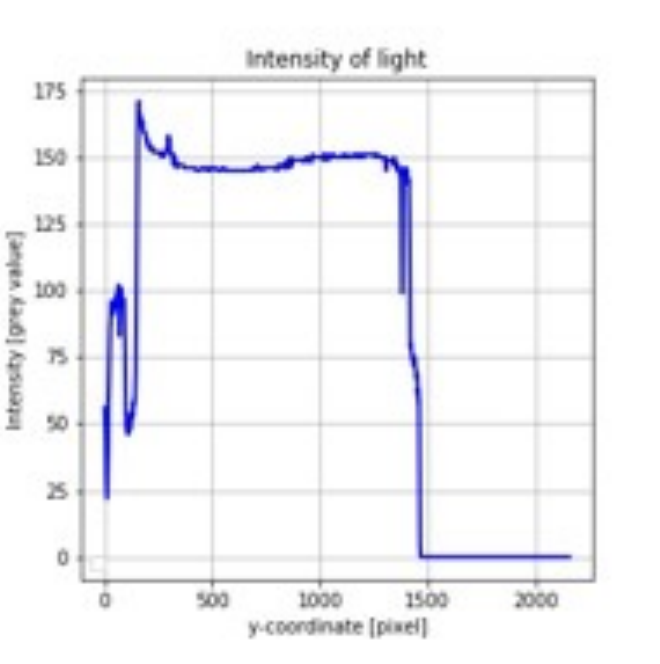}
        \includegraphics[width = 4cm, height = 4cm]{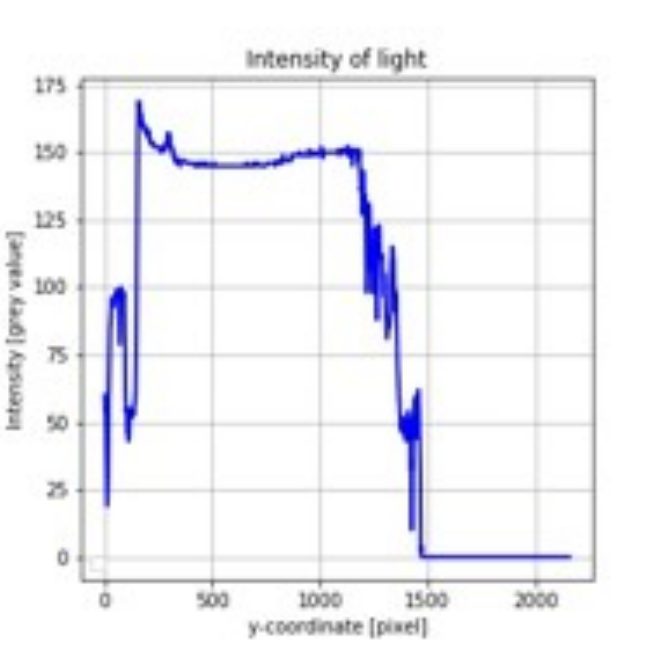}
       
    \caption{ \label{lunframe2}Top: Olivine before (left) and during (right) Lunar gravity. Bottom: greyscale value of light from the LED back panel, which is either transmitted or blocked by the sample. The curve is plotted at fixed X-coordinate, as a function of vertical height. The sharp line around y-values of 1500 represents the transition from volume occupied by densely packed sample, to the empty chamber: Note the visible change in signal at this edge, as the sample barely breaks apart and creates a new (non-sloped) surface during lunar gravity. This phenomena is very consistent with the prediction shown in figure \ref{forceprofile}. }
    \label{lunarframes}
\end{figure}

Having demonstrated the functionality of System 1, we state that the experiment can be used and the results applied in a highly versatile manor. We can, for example, verify scalings to the Klinkenberg correction, in the simplest case under Earth's gravity and with idealized samples, since one-dimensional models of flow through granular material are generally useful and convenient. The unique aspect of this facility is the capability to vary the gravitational load, and in so doing we can pursue at least two avenues of research that are of direct relevance for planetesimal formation and evolution. 

One direction of investigation relates to general prescriptions for two-phase flow with relatively high dust to gas ratio. In particular, we would like to investigate the role of inertia in causing pressure gradients in low-Reynold's number flow, since the classical setup usually only considers porosity as the determinant of the pressure differential. In a recent campaign, we realized an experiment exactly for this goal. Our sample material in the first container was a small quantity of high-density material (steel) as reference to the second two chambers which contained a low-density material (silica): in the second chamber we seeded equivalent mass but greater volume as the reference chamber, and in the third chamber we seeded the same volume but much less mass as the reference material. These measurements will be used to test propositions from authors such as \citep{Lin:2017} and \citep{Squire_Hopkins:2018a}, which imagine that dust-drag fluid instabilities involve localized pressure differentials due to the mass-loading of dust.  

The other direction of investigation relates to the flow dynamics on, near, or through an already formed planetesimal or cometesimal. Due to the low gravity of such small bodies, the packing of the material is more porous than it is on Earth. Indeed the density of the particle packing on comets has been used to infer the formation history, since a collisional growth would result in a denser packing than gravitational instability \citep{mohtashim,Wahlberg:2014}. Within the strict gravitational collapse paradigm, we can then add to the knowledge about how gas percolates through the surface layers of comets: this is important for how sublimated gas might seep into the nucleus to form a sintered cohesive crust, or how gas escapes through the regolith and lifts the dust mantle away from the surface to form a coma. Dust production rates in comae remain an open topic which is treated mainly as a free parameter in simulations, and not one which is fully understood from first principles. Furthermore, there are geological features on comet surfaces, such as aeolean ripples or evidence of mass wasting, which imply more significant atmosphere than is generally assumed, and so sub-surface gas release mechanisms remain an important focus. 

\subsection{Experiment 2: Dust Drag}

This flow chamber contains 1-100 microbar of air with flow rate of approx. 0.1 m/s. The chamber is preloaded with a small quantity of dust analogue material. We report steady operational pressures less than 1 mbar.  

We calculate the terminal velocity, stopping time, and stopping distance, of particles, when the gravitational acceleration is allowed to change. The terminal velocity using the Stokes drag force is $v_{Stokes}=\frac{1}{18}g d_{p}^2\rho /\eta$, and apply the Cunningham correction, $C_{Kn}=1.0+Kn(\alpha+\beta e^{-\gamma/Kn})$, so that the terminal velocity depends upon Kn and is $v(Kn)=C_{Kn}v_{Stokes}$. Here $g=9.8$ m s$^{-2}$ is Earth's gravitational acceleration, and $\eta$ is the molecular viscosity of air. $\alpha$, $\beta$, and $\gamma$ are taken from the experimental literature \citep{Allen_Raabe:1985}. 
   
Since we are interested in order of magnitude estimates, we do not account for porosity, and consider three regimes in material density, that of ice ($\sim$1000 kg m$^{-3}$; solid lines), rock ($\sim$3000 kg m$^{-3}$; dashed lines), and metal ($\sim$8000 kg m$^{-3}$;dot-dashed lines). While our facility is not temperature controlled for the use of ice particles, we have the capability to produce uniquely shaped particles using a micro 3D-printing technique. The ``photoresist" out of which the particles are made has density ~1200 kg m$^{-3}$, and so such particles have close to the same density as ice does and lower density than that which rock has. 

We now express quantities in terms of the friction time $T_{f}$, which is the ratio of the particle momentum to its drag force, and determines the scale time over which the particle reaches its terminal velocity, $v(Kn)=T_{f}g$. The associated distance scale over which a particle reaches its steady state velocity is the friction length, $L(Kn)=T_{f}^{2}g$.

\begin{figure}[ht]
\centering
\includegraphics[width=0.55 \textwidth]{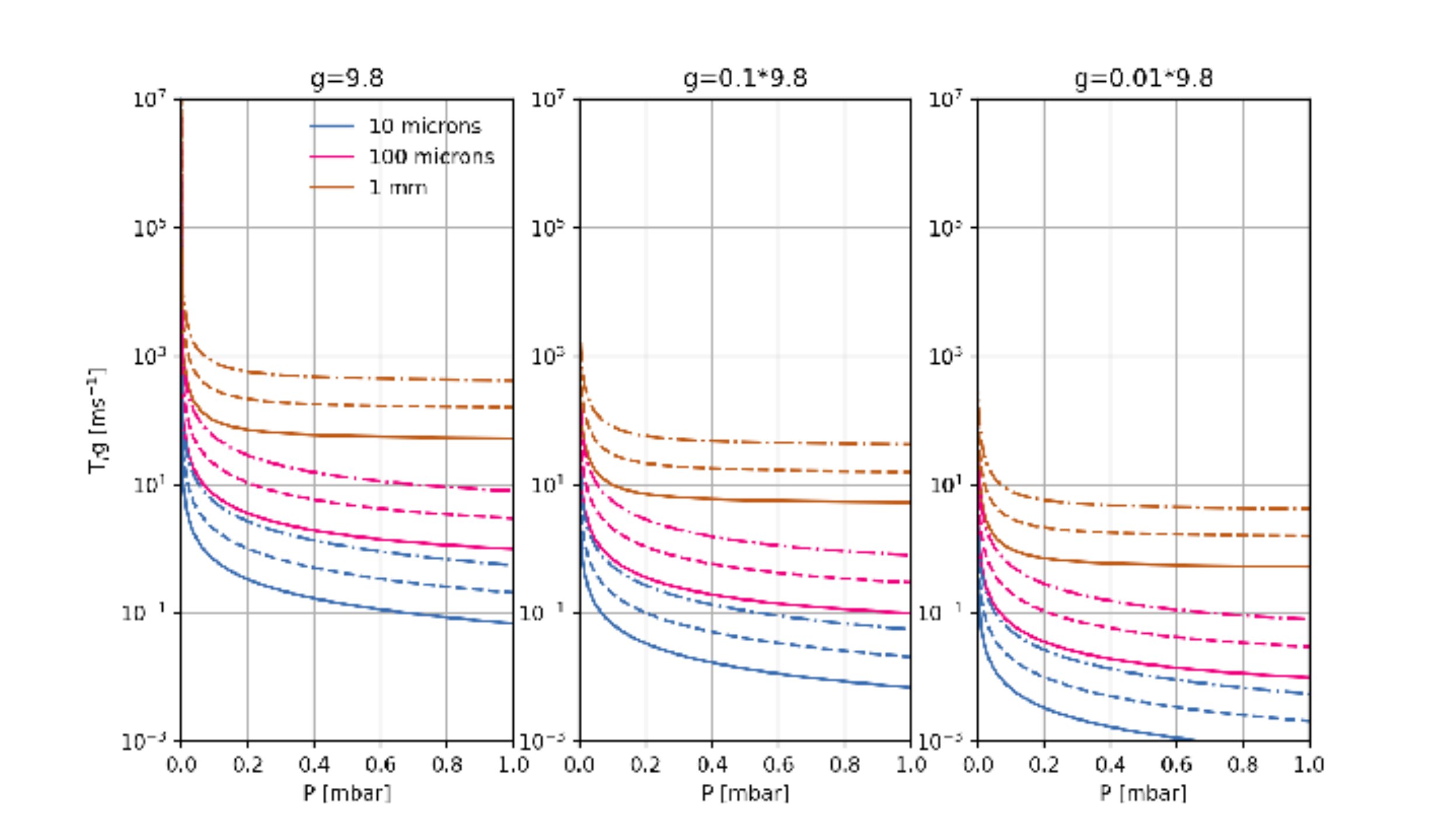}
\caption{\label{friction_time_g} The terminal velocity of a particle, as a function of gas pressure. From left to right: decreasing gravitational load. Solid lines: particles with density of ice; dashed lines: particles with density of rock; dash-dotted lines: particles with density of metal. The blue, fuchsia and brown colours correspond to 1 mm, 100$\mu$m and 10$\mu$m size particles respectively. By reducing gravitational load, we can shorten the time for particles to decelerate in their approach to terminal velocity when traveling in a gas stream.}
\end{figure}

\begin{figure}[ht]
\centering
\label{friction time}
\includegraphics[width=0.55 \textwidth]{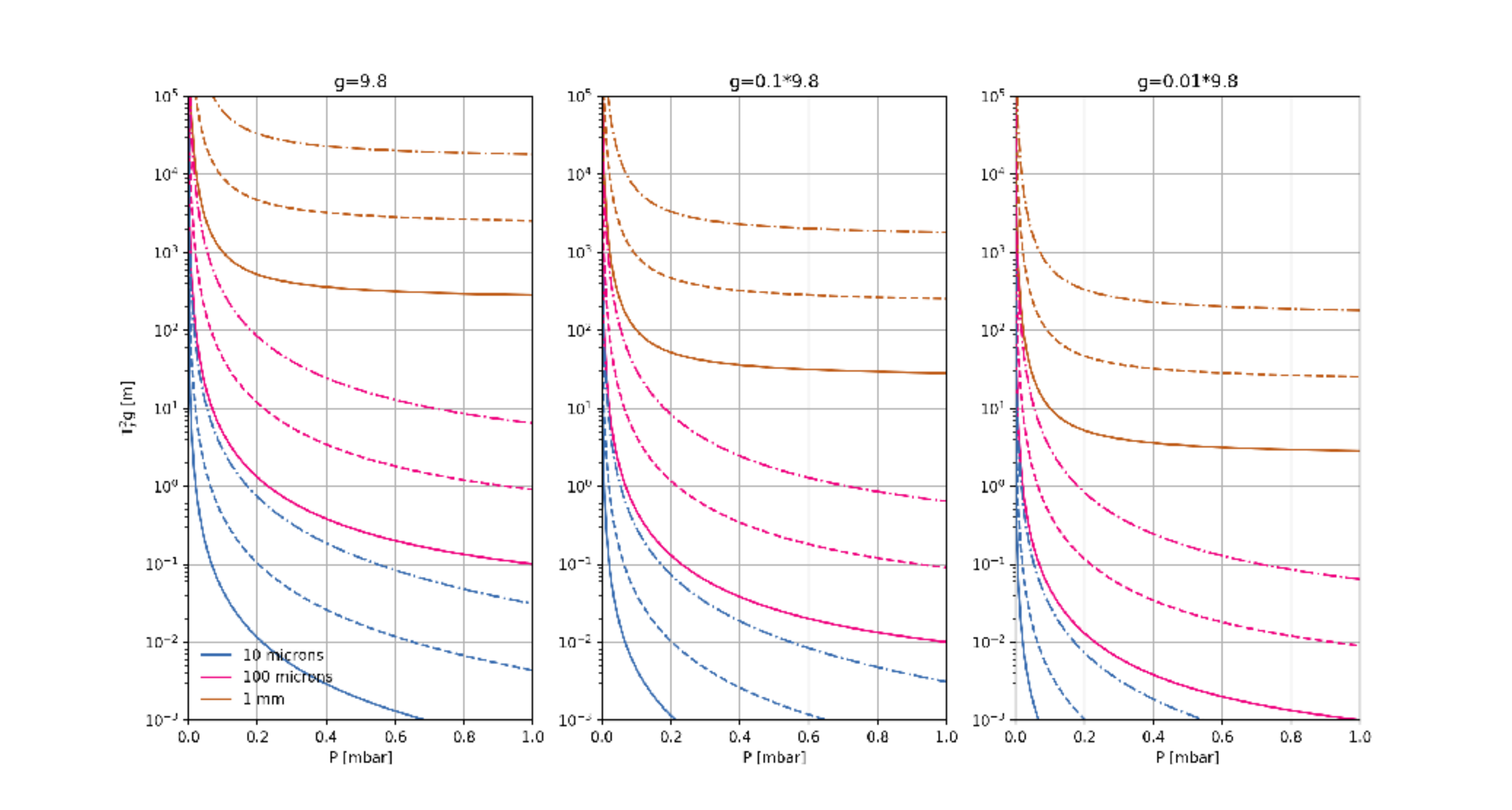}
\caption{The length scale factor over which a particle couples to a gas, as a function of gas pressure. From left to right: decreasing gravitational load. Solid lines: particles with density of ice; dashed lines: particles with density of rock; dash-dotted lines: particles with density of metal. The blue, fuchsia and brown colours correspond to 1 mm, 100 $\mu$m and 10$\mu$m size particles respectively. By reducing gravitational load, we shorten the distance over which particles decelerate in their approach to terminal velocity when traveling in a gas stream.}
\end{figure}

Evidently, particles in the 10-100 micron range have extremely high terminal velocities (several 10s of m/s) when they are approaching drag in the free-molecular flow regime, according to the above, around 0.1-0.5 mbar. However, since the terminal velocity scales directly with g, as the gravitational force decreases, so does the velocity of the particle. We see that for 1/100 the gravitational acceleration of Earth, particles under free-molecular flow conditions may be moving centimeters per second. 

Particles under Earth's gravitational acceleration have stopping lengths covering a distance of several 10s of meters and therefore are not feasible to control or measure over this distance. However, for 1/10 Earth's gravity, 100-$\mu$m ice (or photoresist analogue) particles will stop accelerating over a distance roughly 10 cm, and for 1/100 Earth's gravity, they decelerate within centimeters. 

For particles to be transported across the window with 4 cm diameter, it is required that the speed of the gas is equal to or greater than the terminal velocity of the particles. We have tested the facility with particles of size $500\mu$m and gas pressures $10^{-2}$ mbar. We see from Figure \ref{friction_time_g}, that a relatively low flow speed of $\sim 0.1$ms$^{-1}$ can lift the particles when gravity is reduced. 

We verify the frame crossing time: at a measurement speed of 10 KHz, the particles traverse the window over the course of approximately 10 frames. This gives a velocity of about 2.5 cm s$^{-1}$, in agreement with the estimates. 

A particular feature of dust aggregates in a protoplanetary disc is that they experience a significant relative velocity with respect to the gas velocity. In the planetary sciences literature this is called `headwind' and in basic flow literature, it is often called `slip' velocity; in other words how much the particles move through the fluid, as opposed to simply moving at the same speed as the fluid. We consider particles with high slip to be inertial particles and those with no slip to be tracer particles. It has been found in numerical simulations that both inertial and tracer particles can lead to fluid instabilities in protoplanetary discs. However, their characteristics -- such as typical growth rates, outcomes for dust diffusion, strength of density and velocity gradients in resulting turbulence -- can be quite different. The question of whether a particle is a tracer or an inertial particle is usually determined by {\bf Kn} and the particle mass and is characterized by the Stokes number {\bf St}, which depends directly on drag coefficient. With the experiments in chamber 2, we intend to extract drag coefficients of porous aggregates with high slip velocity. In the event that the drag coefficient is reduced for porous aggregates compared to solid particles of the same dimensions, it would challenge the prevailing understanding of the growth sequence from dust to planetesimals, since the fluid instabilities supposed to aid the formation of planetesimals depend strongly on {\bf St}.

\subsection{Experiment 3: Shear Flow}
To estimate the size and time scales of the instability that may develop at the interface between a dust-laden and dust-free gas, we adapt the result from the linear stability analysis of the Kelvin-Helmholtz instability, which has the following dispersion relation: 

The real part, 
\begin{equation} \label{realpart}
\omega_r/k_{x}=\frac{\rho_{1}v_{1}+\rho_{2}v_{2}}{\rho_{1}+\rho_{2}}
\end{equation}
and the quadratic term has a positive growth rate as long as there is a velocity difference between the two phases: 
\begin{equation}\label{oscillatory}
\sigma/k_{x}=(v_{1}-v_{2})\frac{(\rho_{1}\rho_{2})^{1/2}}{\rho_{1}+\rho_{2}}.
\end{equation}

In equations \ref{realpart} and \ref{oscillatory}, $\rho$ represents density and v represents velocity, with subscripts 1 and 2 corresponding to either the gas or the gas-dust mixture, respectively. 
This gives us a simple way to estimate the wavelengths at which the growth is fastest. However, we note that the velocity and the density are coupled, since in the absence of gravity, the only way to slow the gas phase by the particles is by the collective effect of the mass-loaded two-phase flow. So we first estimate the extent to which the flow can be slowed by a particle phase of a given concentration by evolving the two-fluid equations without external force (and for now, neglecting viscosity). To do so, we use the Euler approach in one dimension, that is, approximating the change in velocity of either the particles or the gas at each time step by summing the derivative. So that the gas velocity, initially 1 ms$^{-1}$, evolves for each time $t_{i}$ as:

\begin{equation}
v_{gas}(t_{i+1})= v_{gas}(t_{i})+\frac{\epsilon}{t_{f}}[v_{dust}(t_{i})-v_{gas}(t_{i})]dt
\end{equation}
and the particle velocity, initially with 0 ms$^{-1}$, evolves for each time $t_{i}$ as:
\begin{equation}
v_{dust}(t_{i+1})= v_{gas}(t_{i})-\frac{1}{t_{f}}[v_{dust}(t_{i})-v_{gas}(t_{i})]dt.
\end{equation}

The velocities of these two coupled phases are essentially symmetric, except the gas phase is controlled by the dust to gas ratio, $\epsilon=\phi \rho_{p}/\rho_{g}$. Note that the filling factor $\phi$ depends strongly on particle size, as $\phi=\pi/6.0 N d_{p}^{3}$, where N is the number of particles of given size per unit volume (assumed to be 1 million, here $d_{p}$ assumed to be 15 micrometers). The densities of the particles are that of rock, 3000 Kg m$^{-3}$, and the gas density can be varied to change $\epsilon$.

\begin{figure}[ht]
    \centering  
        \includegraphics[width = 0.5\textwidth]{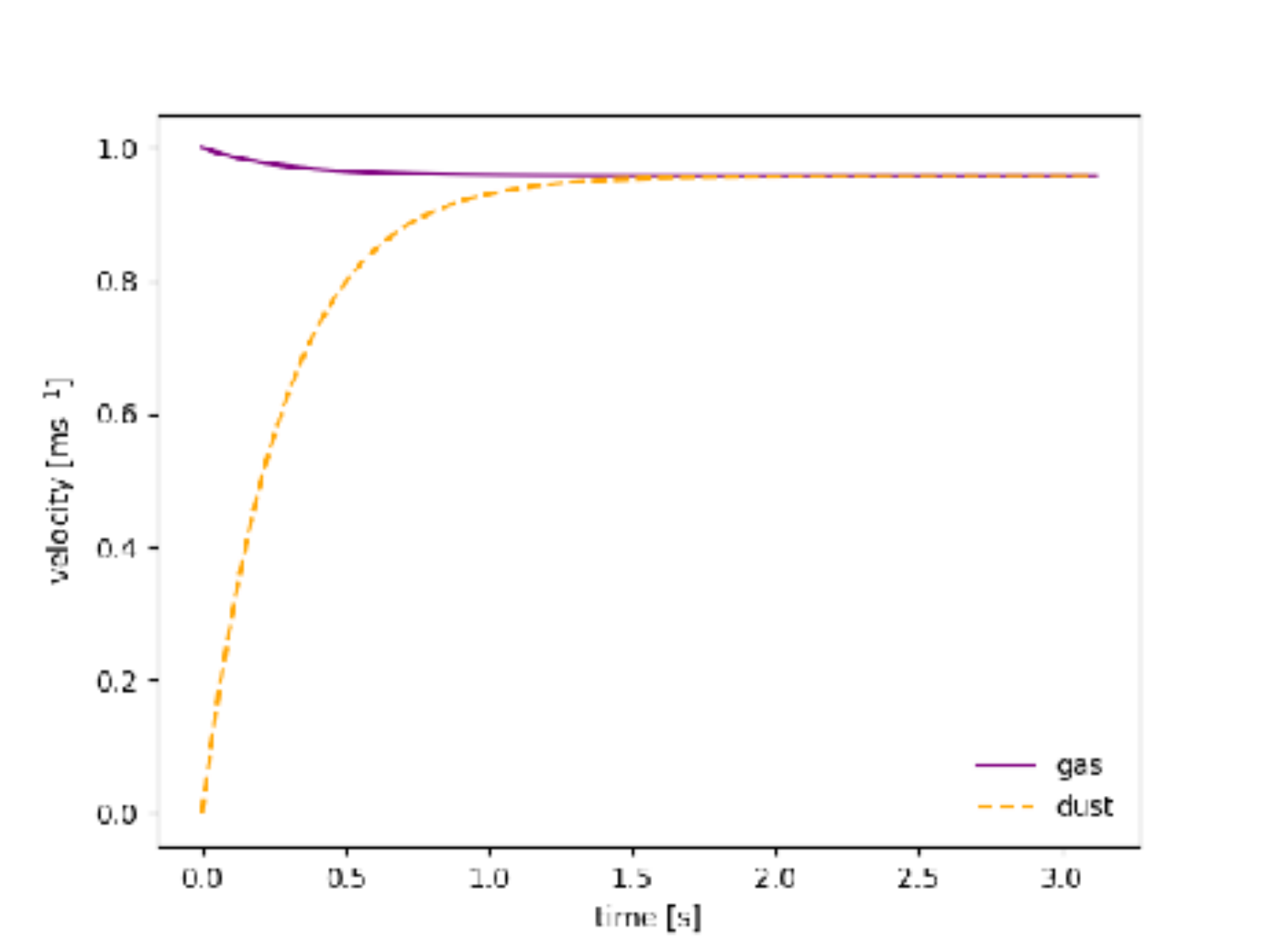}
            \caption{fuchsia solid line represents the gas velocity; dashed orange line represents the dust velocity. The dust, initially at rest, is transported by the gas. Due to the back reaction force, the gas velocity is slowed to converge with the dust velocity, in the layer of flow that is seeded with particles. From Equations \ref{vgas_equation} and \ref{vdust_equation}, it is clear that either a large stopping time, or small dust-to-gas ratio will make the coupling between the two phases too weak for the backreaction force to be important. For 10 mbar gas, initially at 1 m/s, and a mass loading of 0.1, it is possible to slow the gas by $\sim 5-10 \%$.  This will result in a difference in phase velocity, so that, according to Equation \ref{oscillatory}, an oscillatory instability will develop. }
    \label{sys3_dustspeed}
    
\end{figure}

\begin{figure}[ht]
    \centering  
        \includegraphics[width = 0.5 \textwidth]{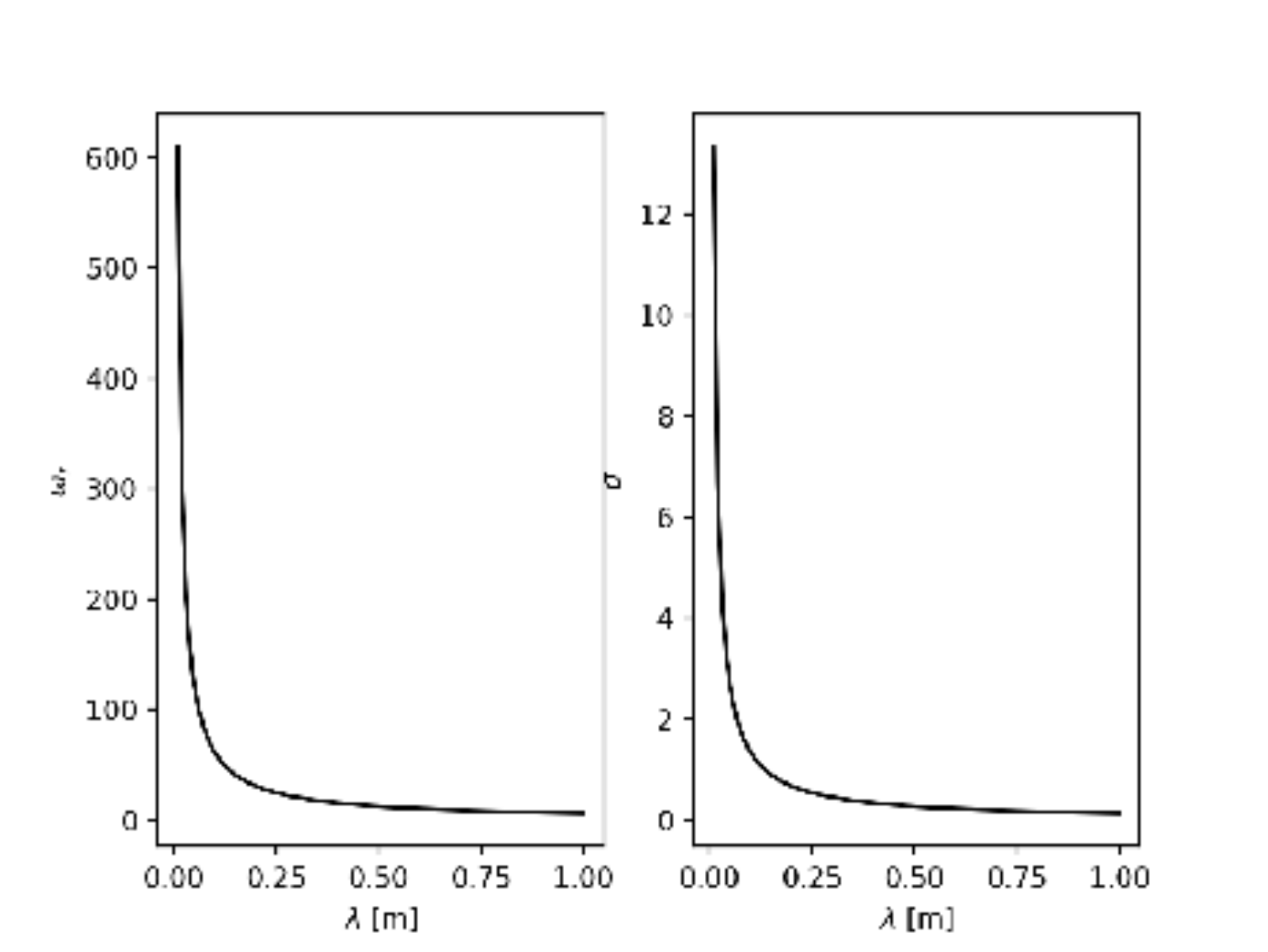}
            \caption{ The real part of the frequency (left) and growth rate (right) of a shear instability as a function of wavelength. Modes with the highest wave numbers will develop the fastest. The wavelength of the fastest growing modes is only a few cm, and so the features that develop in the flow should be resolved within a field of view on this scale.}
    \label{sys3_growth_rate}
    
\end{figure}

Figure \ref{sys3_dustspeed} shows the expectation that, within $e$ friction times, the gas accelerates the particles, and the particles decelerate the gas, so that they approach the same speed, which is about 90\% of the original gas velocity. In this case, the mass loading is 0.05 and the stopping time is 0.30 seconds. We do not expect (nor want) the gas to become streaming unstable, since this would require a value of $\epsilon\ge 0.3$. We also expect the flow to travel just under 30 cm before the particle-laden phase has a significant relative velocity with respect to the adjacent particle-free phase. Taking the long-time value of the gas \& dust velocity from Figure \ref{sys3_dustspeed}, and equation \ref{oscillatory}, we calculate the development of the fastest growing wavenumbers, which are related to the wavelength by $k_{x}=\frac{2\pi}{\lambda}$. 

Both the real (left side) and imaginary (right side) parts of the dispersion relation shown in figure \ref{sys3_growth_rate} are dominated by the wavenumber dependence. We see that the highest frequency wave modes have the fastest growth rates and correspond to the shortest wavelengths. We therefore expect that we can generate and resolve with our measurements the smaller structures -- on the mm-cm length scale -- that result from a flow instability.  Note that there being a relative velocity depends upon the mass-loading and this is set by the pressure. By raising the pressure to 100 mbar, we would lower the mass-loading and therefore a relative velocity would not develop. However, if we go too low in pressure, the stopping time of the particles becomes high and it would take too long for the relative velocity to develop. We have tested the system at a steady pressure of 100$\pm 2$ Pa. 

\begin{figure}[ht]
\includegraphics[width=0.55\textwidth]{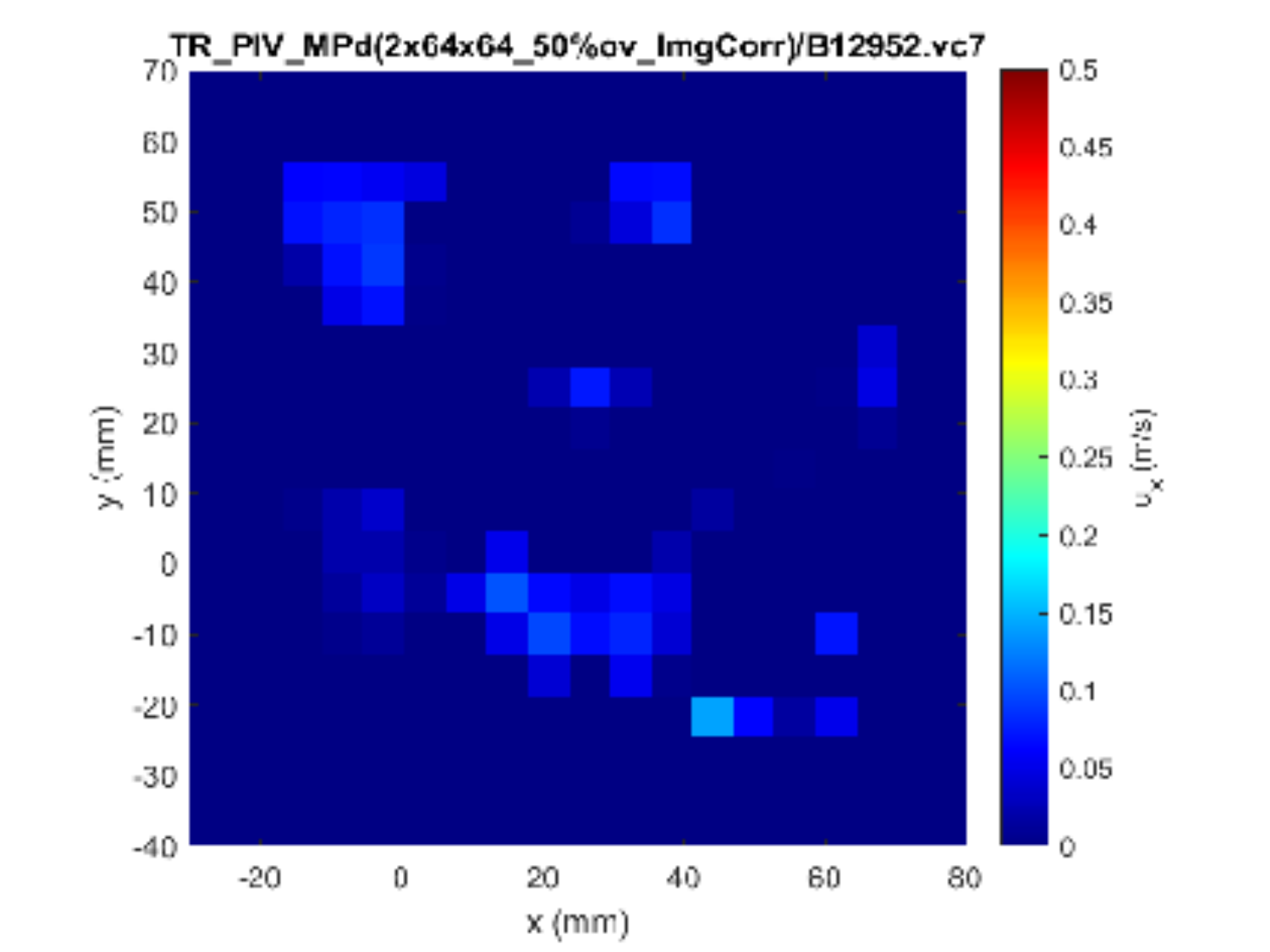}
\caption{\label{vectors} Horizontal (X-direction) component of the velocity field, colored on a scale of 0-0.5 ms$^{-1}$. A typical velocity is 0.1-0.2 ms$^{-1}$, in agreement with Figure \ref{sys3_dustspeed}, which indicates that within the first 30 cm  of travel (location of upstream measurement window), the particles should advance from 0-0.4 m/s in the horizontal direction. The regions with low-velocity voids are expected and characteristic of a pattern formation process.}
\end{figure}

We show a contour plot of the measured horizontal component of the velocity field in Figure \ref{vectors}. The scale ranges from 0-0.5 ms$^{-1}$, with the most typical speed $\sim$0.2ms$^{-1}$. Although Figure \ref{sys3_dustspeed} indicates that the relaxed velocity of the dust and gas should be closer to 0.9 ms$^{-1}$, it is also clear that the dust can be transported up to a half meter before reaching such speed. Since the upstream measurement point is about 0.3 m from the dust injection point, it is logical that the dust has not yet had time to reach its maximum velocity at this location. Nevertheless, there is already a strong relative velocity between the two phases. We also note the inhomogeneous velocity field pattern in Figure \ref{vectors}: the time series (not shown) reveals that this spotted pattern is not random, but displays an interesting periodicity. A separate letter exploring the complex fluid dynamics is in preparation (Capelo et al 2022B). 

To motivate the TEMPus VoLA facility, we introduced the concept of dust-drag fluid instability. There are now a broad class of proposed fluid instabilities, leading to turbulence, which arise principally due to differential motion between the gas and dust phases of two-phase flow, which have been given the name resonant drag instability (RDI). The important quantities that govern the flow behavior are the Mach number, $\epsilon$, and {\bf St}, as well as which ever external forces are present. The streaming Instability originally proposed by \cite{You_good2005} is considered a particular case of RDI. Similarly, the disc sedimentation instabilities proposed by \cite{lambrechts} and \citep{Squire_Hopkins:2018a}, and demonstrated experimentally by \cite{Capelo:2019}, show a universal tendency for such class of fluid instabilities to produce a solid concentration effect, purely via aerodynamical means. The way by which solids form high-density regions is critical for the planet formation process, the understanding of which has long been plagued by the problem that turbulence arising primarily in the gas phase can serve to disperse and mix the solid phase, and prevent favoured gravitational fragmentation scenarios such as proposed by \cite{Safronov:1969} or \cite{goldreich_ward}. 

The experiments in system 3 speak most directly to the type of shear instability supposed to challenge both the Goldreich-ward mechanism and possibility the Streaming Instability: namely disc-plane Kelvin Helmholtz instability. Proponents of the Streaming Instability as the pathway to planetesimal formation conclude that despite the competing effect of Kelvin-Helmholtz instability, the requisite value for streaming instability of $\epsilon=1$ can still be reached \citep{Johansen2007,bai_stone1}. It is worth noting that both the disc-midplane instability and the streaming instability are instances where large-scale coriolis forces should be at play. However, having established the underlying mechanisms essential to these phenomena -- differential dust-gas motion driven by a pressure gradient in one case, and shear between layers of high- and low- solid density in another -- we can proceed to experimentally test fluid behavior in general, when these basic requirements are met. The very general smooth power-law disc model has been challenged in recent years by observations of significant dust structures in protoplanetary discs, which are assumed to host the planet formation process. Even within this framework, localized pressure-fronts with dust pileups develop, and this too can generate a shearing-type instability \citep{Surville:2018}.

All arguments about flow instabilities in protoplanetary discs pre-suppose that the models used in simulations of dust and gas capture the particle behavior correctly. This assumption can be brought into question by the fact that two-fluid models are approximate, in that the coupled equations evolve only volume-averaged quantities of dust density. Moreover, there exist some examples of granular shearing type instabilities, however usually when the grains are met with atmospheric conditions on Earth, such as desert dune ripples or ash in volcanic plumes. However, no known examples of shear instability with rarefied gas has ever been generated.  In experiment 3, we specifically look at the shearing process, first to demonstrate whether an instability exists at all. Then, the capability to systematically change the gas pressure in the system and to change the particle types allows us to vary both $\epsilon$ and $T_{\rm f}$. In principle, we should cross a regime transition where the particles experience greater and greater slip, and either we have only a RDI, only a shear instability, or the co-existence of both. We can also address more subtle questions about realistic particle velocity distributions in unstable flow -- with implications for collision frequency and therefore aggregate growth. Another subtlety is the treatment of viscosity, which has been predicted to change in the presence of particles \citep{einstein_viscosity:1905}, which is important to know, since viscosity sets the dissipation length scale of turbulent flow.

\section{Conclusions}\label{future}
\subsection{Summary}
We presented the results of the design and testing of a facility that contains three integrated, but independent, experiments. Each experiment explores a different aspect of particle-gas interaction in the very dilute gas context. The physical processes we investigate are analogous to the processes involved in the formation of planetesimals and their evolution as comets or asteroids. In this manuscript we showed data from all three experiments, confirming that the experiments are working within their design concept, that the hardware is sound, and that the measurement acquisition methodologies are robust and reliable. Moreover, our data confirm that it is possible to access the parameter range that is relevant for applications to the planet-formation scenario. 

The data presented was acquired over the course of two parabolic flight campaigns aboard the aircraft Air Zero-G, operated by Novespace. The entire facility was first commissioned in the fourth Swiss PFC, consisting of 16 parabolas organised by the University of Zurich Space Hub.  During this campaign, all of the constructed experimental apparatus were installed in the racks. The value of this pilot flight was to certify the safety of our experiment and gain experience operating the control and measurement systems in flight. We also obtained useful data from the first experiment on gas permeability during this campaign, for example see Figure \ref{pressure} and Figure \ref{lunframe2}. 

We flew the facility again during the 75th European Space Agency (ESA) PFC, which consisted of 93 parabolas over a three-day campaign. We operated all three experiments during this campaign. The facility is scheduled to be flown on the 78th ESA PFC, during which we will extend the suite of measurements to include a wider variety of particle types and operating conditions.  

\subsection{Outlook}

The three limiting cases represent a full exploration of the Darcy-Brinkman momentum and
mass conservation equations\citep{Brinkman} which presents the formal equivalence
between granular gas permeability and tortuosity drag force. These formulae have historically
been applied to granular two-phase flows, but remain unexplored in both the Epstein drag
regime and in the limit of ultra-high porosity, and therefore were previously of limited use in
an astrophysical context. An important subtlety of this theory is that it prescribes a
modification to the fluid viscosity due to the presence of the particulate phase \citep{einstein_viscosity:1905}.
Given the central -- however elusive -- role of viscosity in driving turbulence, accretion, and
structure formation in the context of star and planet formation, the physics under
consideration is of fundamental importance.
A defining characteristic of dust-drag fluid instability proposed to be operating in
protoplanetary discs is that there must be a net relative motion between dust and gas, and that
the collective forcing between particles of a given Stokes number causes the development of
localized pressure gradients \citep{You_good2005,Surville:2018,Capelo:2019}. This study is unique because we are not simply studying gas-particle interaction,
either in vacuum, or under weightlessness conditions: we specifically drive a pressure
gradient of low-pressure gas in order to cause fluidization and transport of the dust-analogue
material. We utilize the variation in gravity throughout the parabola to change the inertia of
the particles and hence vary the relative particle-gas velocity. In the third system, we mainly
exploit the Zero-G phase, but again, it is by the gradient-driven flow that we transport the
particles and study the behavior of the fluid with when the external body force is low. To our
knowledge, no facility specifically addressing simultaneously these dynamical criteria and
requisite flow conditions has ever been flown in parabolic flights or other microgravity
platforms.

While vacuum chambers are common, spatially resolved studies of fluid flow under vacuum
conditions are challenging and rarely conducted due mostly to technical limitations. In the
fluid dynamics, as well as the automotive and aerospace engineering communities, some of
the most typical measurement techniques used to address complex flow are based upon
seeding `tracer’ particles into the fluid and using multiple high-speed cameras to reconstruct
the tracer positions. Using various analysis tools, one can then derive properties from the
particle-tracking data, such as fluid density, typical velocities and accelerations, shear and
strain rates, etc. Similar methods have been used to study ‘inertial’ particles that respond to
fluid flow. However, rarely have such high-precision methods been
used to study rarefied gas fluid dynamics \citep{Capelo:2019}. Despite the challenging nature
of such measurements, it is of high importance to continue this line of research in order to
access the unique and proper parameter space (high-Knudsen number two-phase flow) that is
applicable directly to fluid instabilities in the planetesimal formation scenario.
While the facility we designed is both ambitious and complex, we have already solved the
most challenging design problems and we expect that over the life of the facility and via
several campaigns, we will provide an unprecedentedly thorough exploration of dust-gas
interaction in a flow regime that is widely studied in theory and simulations of protoplanetary
discs and planetesimal formation and evolution, but as yet practically untouched in the
experimental literature.

In the process of addressing gas permeability, drag force, and shear stress, we will
inherently obtain the tensile and shear strength of our samples. Future experimental
campaigns represent therefore opportunities to vary our samples so as to extend the existing
database of dust tensile \citep{Gundlach:2018} stress measurements to include
additional minerals which must have formed in the earliest epoch of the solar system
and are common to chondritic meteorites and cometary dust. Such data is useful to
improve models of planetesimal evolution due to collisions \citep{Jutzi:2017} and
outflows \citep{Thomas:2015}.

\begin{acknowledgments}
This work has been carried out within the framework of the NCCR PlanetS supported by the Swiss National Science Foundation. JB and RS acknowledge the continuous support by the Deutsches Zentrum für Luft- und Raumfahrt (DLR). The authors recognise the ongoing support of The Swiss Space Office  and the Programme de Développement d'Expériences Scientifiques (PRODEX) of the European Space Agency (ESA). We thank  Prof. Oliver Ulrich for planning the Swiss Parabolic flight program and the University of Zurich Space Hub. This project has benefited from the technical support from Novespace. 
\end{acknowledgments}

\bibliography{methods_hlc.bib}
\appendix*
\section{to-scale-drawings}\label{append}

\clearpage
\pagebreak

\begin{figure}
\includegraphics[angle=90, width =  0.9 \textwidth]{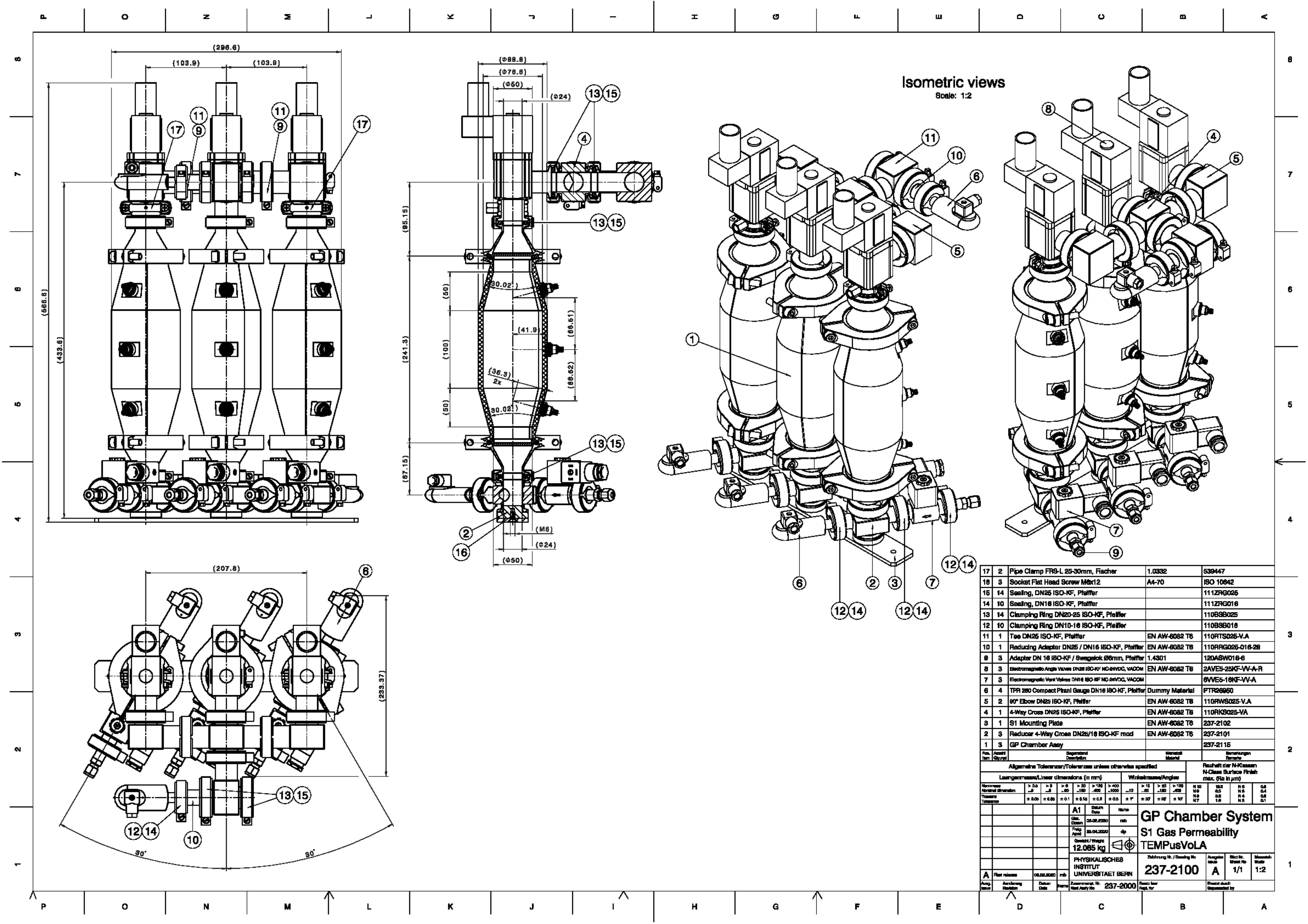}%
\caption{\label{mechanical:exp31} Mechanical drawing and detail view of experiment one, gas permeabilty chamber.}%
\end{figure}

\clearpage
\pagebreak

\begin{figure}
\includegraphics[angle=90,width = 0.9 \textwidth]{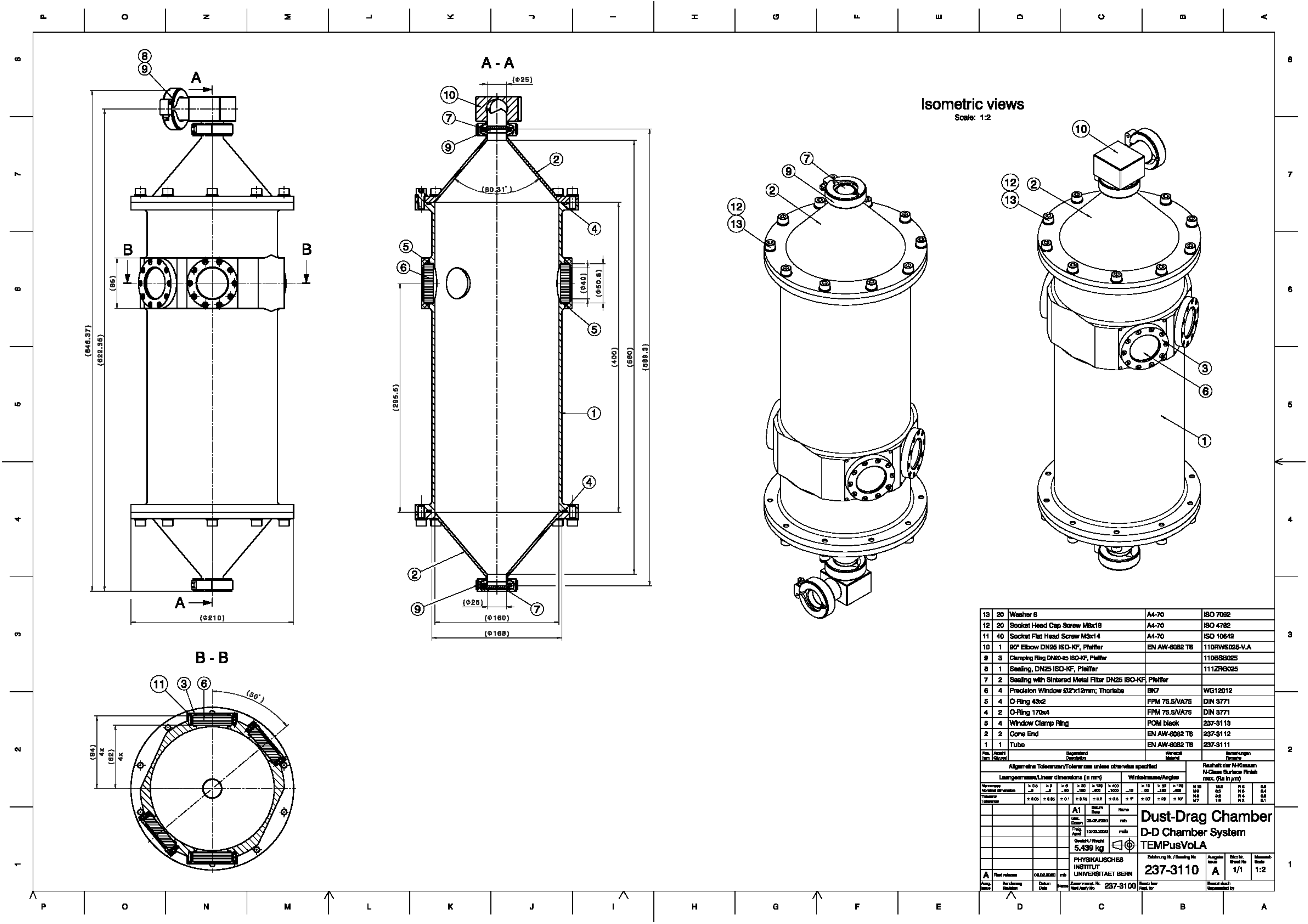}%
\caption{\label{mechanical:exp32}Mechanical drawing and detail view of experiment two, dust drag chamber.}%
\end{figure}
\clearpage
\pagebreak

\begin{figure}
\includegraphics[angle=90,width = 0.9 \textwidth]{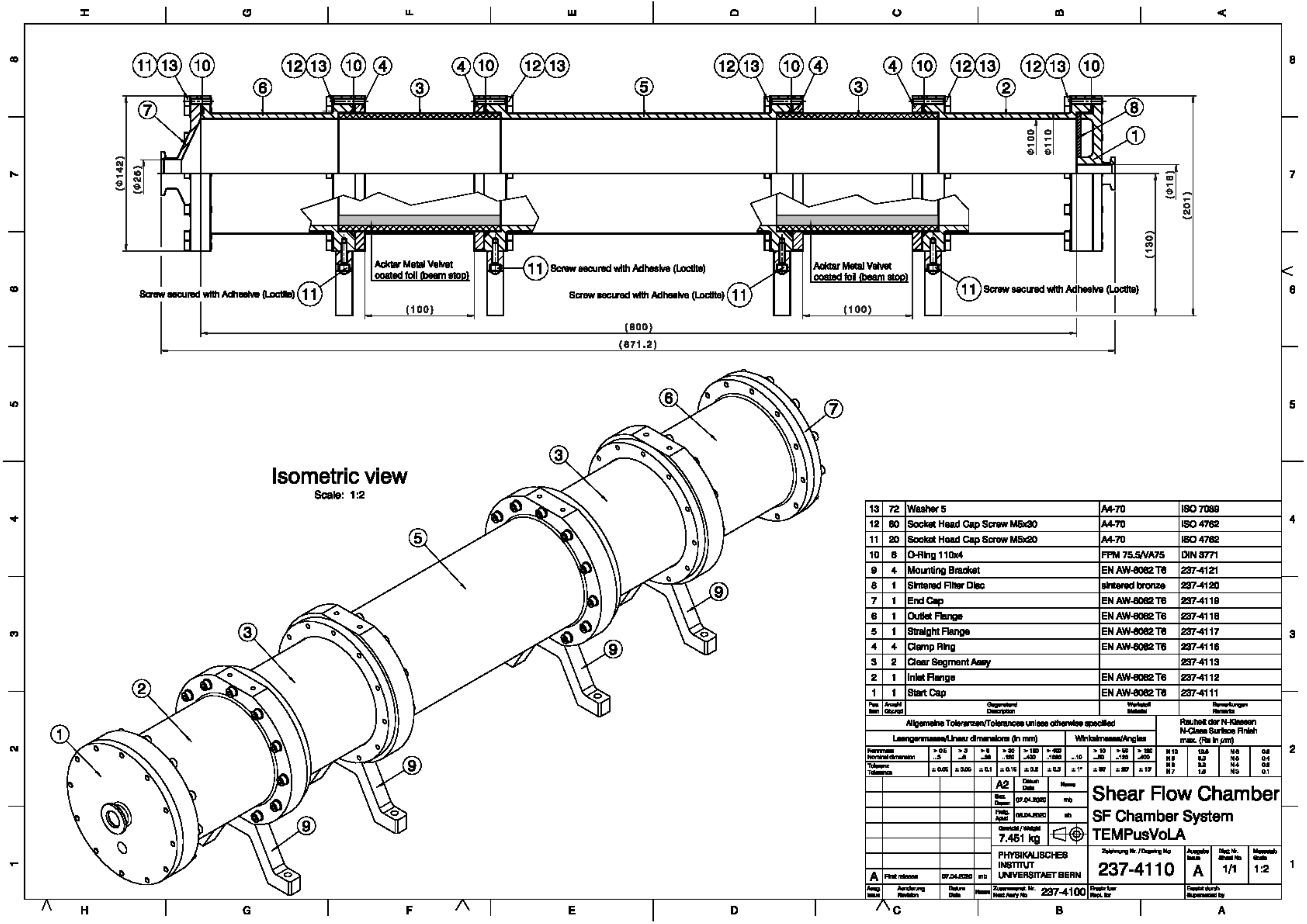}%
\caption{\label{mechanical:exp33} Mechanical drawing and detail view of experiment three, dusty shear flow chamber. }%
\end{figure}


\end{document}